\documentclass[aip,jcp,longbibliography,amssymb,amsmath,reprint,superscriptaddress]{revtex4-2}
\usepackage{graphicx}
\usepackage{bm}
\usepackage{multirow}
\usepackage{color}
\usepackage{braket}
\usepackage{algorithm}
\usepackage{algpseudocode}
\usepackage{setspace}
\usepackage{subcaption}
\usepackage[colorlinks]{hyperref}
\hypersetup{ 
    colorlinks=true,
    linkcolor=red!80!black,
    urlcolor=magenta!80!black,
    citecolor=green!70!black,
}
\usepackage{upgreek}
\usepackage{booktabs}
\usepackage[round-mode=places,round-precision=6]{siunitx}
 \usepackage[T1]{fontenc}
\usepackage{dcolumn}
\usepackage{xcolor}
\usepackage{verbatim}
\usepackage{caption}
\usepackage{xcolor} 

\newcommand*{\citen}[1]{%
  \begingroup
    \romannumeral-`\x 
    \setcitestyle{numbers}%
    \cite{#1}%
  \endgroup   
}
\makeatletter
\newcommand*{\rom}[1]{\expandafter\@slowromancap\romannumeral #1@}
\makeatother

\newcommand{\ee}{\text{e}}
\newcommand{\ii}{\text{i}}
\newcommand{\dd}{\text{d}}
\newcommand{\rrangle}{\rangle\!\rangle}
\newcommand{\llangle}{\langle\!\langle}
\newcommand{\bbrakket}[1]{\llangle #1 \rrangle}

\begin{document}

\title{Adiabatic extraction of nonlinear optical properties from real-time time-dependent electronic-structure theory}

\author{Benedicte Sverdrup Ofstad}
\email{b.s.ofstad@kjemi.uio.no}
\affiliation{Hylleraas Centre for Quantum Molecular Sciences, Department of Chemistry, 
             University of Oslo, Norway}
\author{H{\aa}kon Emil Kristiansen}
\affiliation{Hylleraas Centre for Quantum Molecular Sciences, Department of Chemistry, 
             University of Oslo, Norway}
\author{Einar Aurbakken}
\affiliation{Hylleraas Centre for Quantum Molecular Sciences, Department of Chemistry, 
             University of Oslo, Norway}
\author{{\O}yvind Sigmundson Sch{\o}yen}
\affiliation{Department of Physics, University of Oslo, Norway}
\author{Simen Kvaal}
\affiliation{Hylleraas Centre for Quantum Molecular Sciences, Department of Chemistry, 
             University of Oslo, Norway}           
\author{Thomas Bondo Pedersen}
\email{t.b.pedersen@kjemi.uio.no}
\affiliation{Hylleraas Centre for Quantum Molecular Sciences, Department of Chemistry, 
             University of Oslo, Norway}

\date{\today}

\begin{abstract}
Real-time simulations of laser-driven electron dynamics contain information about molecular optical properties through all orders in response theory.
These properties can be extracted by assuming convergence of the power series expansion of induced electric and magnetic multipole moments.
However, the accuracy relative to analytical results from response theory quickly deteriorates for higher-order responses
due to the presence of high-frequency oscillations in the induced multipole moment in the time domain.
This problem has been ascribed to missing higher-order corrections. We here demonstrate that the deviations are caused by nonadiabatic effects arising from
the finite-time ramping from zero to full strength of the external laser field.
Three different approaches, two using a ramped wave and one using a pulsed wave, for extracting electrical properties from real-time time-dependent electronic-structure simulations are investigated.
The standard linear ramp is compared to a quadratic ramp, which is found to yield highly accurate results for polarizabilities, and first and 
second hyperpolarizabilities, at roughly half the computational cost. 
Results for the third hyperpolarizability are presented along with a simple, computable measure of reliability.
\end{abstract}

\maketitle

\section{Introduction}

Extraction of frequency-dependent, off-resonance linear and nonlinear optical properties of molecules
from real-time time-dependent electronic-structure simulations
has been increasingly
used~\cite{Yabana1996,Tsolakidis2002,Yabana2006,Wang2007,Wang2007a,ding,yamaguchi_large_2016,konecny_acceleration_2016,yatsui_nano_scale_2017,lestrange_chapter_2018,goings_real_time_2018,PW,li_real-time_2020,baiardi_electron_2021,omp2,omp2_cor}
in place of conventional response theory~\cite{Olsen1985,Christiansen1998b} in recent years.
One likely reason is that the implementation of response theory becomes increasingly cumbersome with increasing response order,
whereas time-dependent methods are relatively straightforward to implement.

While response theory is based on perturbation expansions, real-time approaches where the initial ground-state
wave function (or density or density matrix)
is propagated in the presence of an external laser field automatically include responses to all orders in perturbation theory.
In principle, therefore, optical response properties through \emph{any} order can be extracted from induced multipole moments recorded during
the real-time simulation. If the total number of time steps in the wave function propagation can be kept low enough,
the real-time approach may become computationally advantageous over the response approach for higher-order nonlinear properties
such as the second hyperpolarizability.

The time required for real-time simulations depends on several parameters besides the inherent computational complexity of the equations of motion for the wave function parameters, which generally take the form $\dot{y} = f(y, t)$ where the dot denotes the time derivative.
The choice of integrator affects how large a time step may be used without sacrificing accuracy and
the number of expensive evaluations of the function $f(y,t)$ per time step. For given choices of electronic-structure model and
suitable integrator, however, the key parameter determining both computational effort and accuracy of the extracted properties
is the form used for the external, time-dependent field.

Two general approaches for the extraction of response properties have been proposed recently.
While \citeauthor{PW}~\cite{PW} used a pulsed (i.e., with finite duration) wave,
\citeauthor{ding}~\cite{ding} used a monochromatic continuous wave ramped from zero to full strength in a finite-time interval to
mimic the adiabatic switching-on required by response theory.
In both cases, the individual orders of the response of the electronic system are separated by running simulations with 
different field strengths, followed by curve fitting to extract specific frequency-dependent properties at each order.
The accuracy and total simulation time thus intrinsically depend on the duration of the pulsed or continuous wave, including the
ramping time for the latter.

In this work, we investigate the convergence of the extracted response properties (polarizabilities, and first and second hyperpolarizabilities)
towards the results from response theory with respect to the duration of the pulsed wave (PW). 
For the ramped continuous wave (RCW) approach, we perform the same convergence study with respect to 
the adiabatic ramping time and the post-ramp time.

\begin{figure}[h]
    \subfloat{{\includegraphics[scale=0.65]{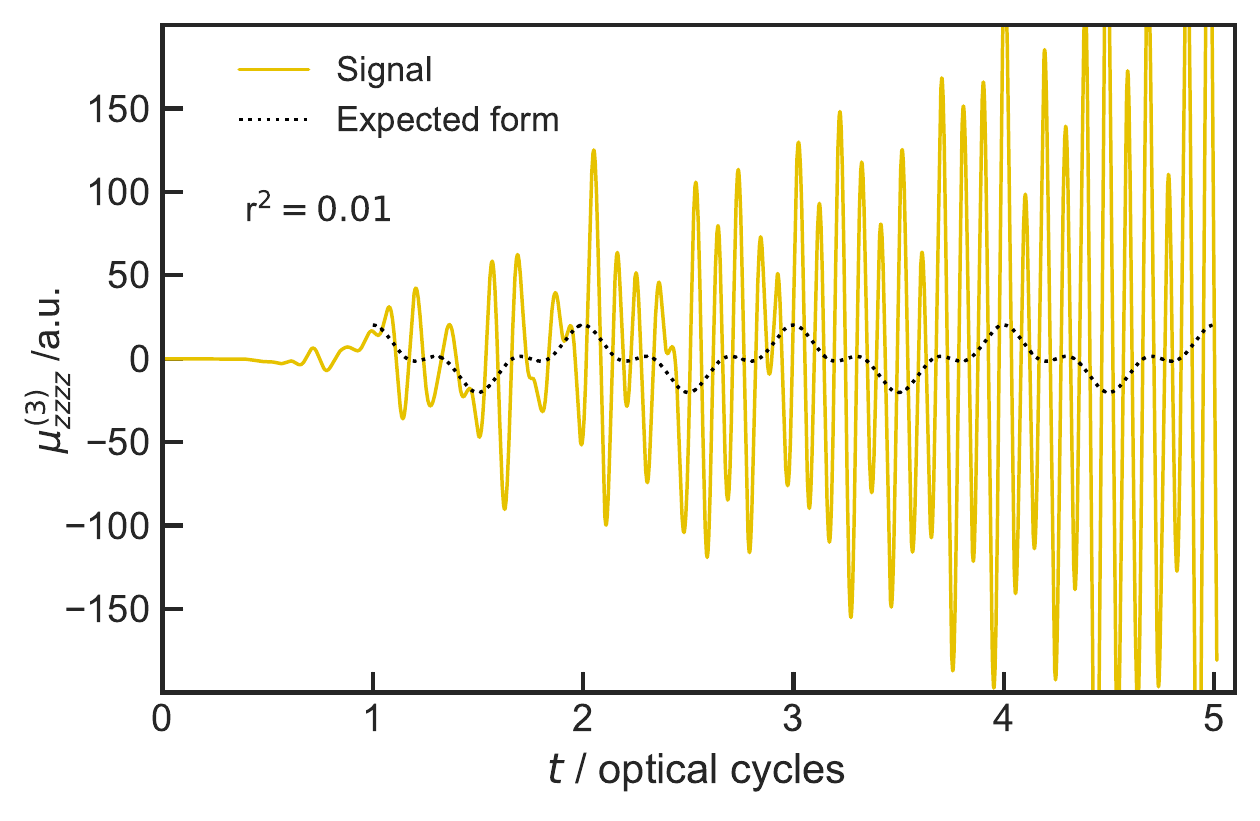} }}
    \caption{Comparison of the third-order induced dipole moment extracted from TDCCSD simulations of the Ne atom with a least-squares fitting to
    the form expected from response theory.\label{intro_fig}}
\end{figure}

The results reported by \citeauthor{ding}~\cite{ding} indicate that
linear polarizabilities can be extracted from simulations with errors below $\sim 1\%$, while the percentwise errors increase by
roughly a factor of ten at each nonlinear order for hyperpolarizabilities. The source of these errors is the significant deviation of the
higher-order time-domain dipole signals from the form expected from frequency-dependent response theory.
Fig. \ref{intro_fig} shows an example for the Ne atom where we have
extracted the third-order induced dipole moment from time-dependent coupled-cluster singles-and-doubles (TDCCSD) simulations
with the approach recommended by \citeauthor{ding}~\cite{ding}
That is, we have used a continuous wave linearly ramped for one optical cycle, followed by wave-function propagation with full field strength for 
four optical cycles. Evidently, the deviation between the computed dipole and the fitted one is much too big for an accurate determination of
the second hyperpolarizability, as also indicated by the coefficient of determination, $r^2=0.01$.

Such deviations have been ascribed to higher-order truncation errors.~\cite{ding}
In this work we investigate if the deviations can be reduced by switching to a softer adiabatic ramping, while maintaining or reducing the
total computational cost of the time-dependent simulations.

The remainder of the paper is organized as follows. 
In Sec.~\ref{sec:theory} we review the RCW and PW methods for extracting up to the third hyperpolarizability and propose an alternative
to the linear ramping of \citeauthor{ding}~\cite{ding} aimed at mitigating nonadiabatic effects.
Test systems and other computational details are provided in Sec.~\ref{sec:compdet}, followed by presentation and discussion of results
in Sec.~\ref{sec:results}. Finally, concluding remarks are given in Sec.~\ref{sec:concluding}.

\section{Theory}
\label{sec:theory}

The electronic dynamics induced by an electromagnetic field is governed by
the time-dependent Schr\"{o}dinger equation,
\begin{equation}
\label{TDSE}
 \ii \frac{\partial \Psi(t)}{\partial t} = \hat{H}(t) \Psi(t),  \quad \Psi(0) = \Psi_0,
\end{equation}
where $\Psi_0$ is the initial condition, here chosen to be the normalized, time-independent ground-state wavefunction.
The time-dependent Hamiltonian $\hat{H}(t)$ is given by 
\begin{equation}
\hat{H}(t) = \hat{H}_0 + \hat{V}(t),
\end{equation}
where $\hat{H}_0$ is the molecular electronic Hamiltonian in the clamped-nuclei Born-Oppenheimer approximation, and
the matter-field interaction operator $\hat{V}(t)$ is given in the length-gauge electric-dipole approximation as
\begin{equation}
\hat{V}(t) = -\hat{\mu}  \cdot  E(t).
\end{equation}
Here, $\hat{\mu}$ is the electric-dipole operator and $E(t)$ is a uniform classical electric field.
The time evolution of the electric-dipole moment is obtained from the explicitly time-propagated wavefunction as
\begin{equation}
\mu(t) =  \bra{\Psi(t)} \hat{\mu} \ket{\Psi(t)}.
\end{equation}

Provided the external field is sufficiently weak and adiabatically switched on,
each Cartesian coordinate of the time-dependent dipole moment $\mu_i(t)$ can be expanded as a power series in the electric field $E_j(t)$.
Separating the electric-field component $E_j(t) = E_jF(t)$ into a constant amplitude $E_j$ and a time-dependent
function $F(t),\ \vert F(t) \vert \leq 1$, we may write~\cite{butcher_cotter_1990}
\begin{align}
\mu_i(t) &= \mu_i^0 + \sum_j \mu^{(1)}_{ij}(t) E_j + \sum_{jk} \mu^{(2)}_{ijk}(t) E_j E_k \nonumber \\
         &+ \sum_{jkl} \mu^{(3)}_{ijkl}(t) E_j E_k E_l \nonumber \\
         &+ \sum_{jklm} \mu^{(4)}_{ijklm}(t) E_jE_kE_lE_m + \cdots.
\label{power}
\end{align}
The time-dependent dipole responses $\mu^{(n)}(t)$ can be written either in the time domain or in the frequency domain.

In the time domain, we may write the dipole responses as the convolutions
of time-dependent polarizabilities and hyperpolarizabilities with the field factors $F(t)$,~\cite{butcher_cotter_1990}
\begin{subequations}
\begin{align}
\label{time-domain2_1}
    &\mu^{(1)}_{ij}(t) =  \int_{-\infty}^{\infty} \alpha^{(1)}_{ij}(t-t_1)F(t_1) \dd t_1, \\
\label{time-domain2_2}
    &\mu^{(2)}_{ijk}(t) = \iint_{-\infty}^{\infty}  \alpha^{(2)}_{ijk}(t-t_1, t-t_2)F(t_1)F(t_2) \dd t_1 \dd t_2, \\
\label{time-domain2_3}
    &\mu^{(3)}_{ijkl}(t) =  \iint\!\!\!\int_{-\infty}^{\infty}  \alpha^{(3)}_{ijkl}(t-t_1, t-t_2, t-t_3) \nonumber \\
                     &\qquad\qquad\quad\times F(t_1)F(t_2)F(t_3) \dd t_1 \dd t_2 \dd t_3, \\
\label{time-domain2_4}
    &\mu^{(4)}_{ijklm}(t) =  \iint\!\!\!\iint_{-\infty}^{\infty}  \alpha^{(4)}_{ijklm}(t-t_1, t-t_2, t-t_3, t-t_4) \nonumber \\
                     &\qquad\qquad\quad\times F(t_1)F(t_2)F(t_3)F(t_4) \dd t_1 \dd t_2 \dd t_3 \dd t_4.
\end{align}
\label{time-domain2}%
\end{subequations}
By causality, the time-dependent (hyper-)polarizability tensors $\alpha^{(n)}$ vanish when any of the arguments $t-t_n < 0$.
In the frequency domain, adopting the conventions of response theory,~\cite{Olsen1985}
\begin{subequations}
\begin{align}
    &\mu_{ij}^{(1)}(t) = \int_{-\infty}^{\infty} \alpha_{ij}(-\omega;\omega) \tilde{F}(\omega) \ee^{-\ii\omega t} \dd \omega,
    \\
    &\mu^{(2)}_{ijk}(t) = \frac{1}{2} \iint_{-\infty}^{\infty} \beta_{ijk}(-\omega^{(2)};\omega_1,\omega_2) \nonumber \\
    &\qquad\qquad\quad\times \tilde{F}(\omega_1)\tilde{F}(\omega_2) \ee^{-\ii(\omega_1+\omega_2) t} \dd \omega_1 \dd \omega_2, \\
    &\mu^{(3)}_{ijkl}(t) = \frac{1}{6} \iint\!\!\!\int_{-\infty}^{\infty} \gamma_{ijkl}(-\omega^{(3)};\omega_1,\omega_2,\omega_3) \nonumber \\
    &\qquad\qquad\times \tilde{F}(\omega_1)\tilde{F}(\omega_2)\tilde{F}(\omega_3)
                        \ee^{-\ii(\omega_1+\omega_2+\omega_3) t} \nonumber \\
    &\qquad\qquad\times \dd \omega_1 \dd \omega_2 \dd \omega_3, \\
    &\mu^{(4)}_{ijklm}(t) = \frac{1}{24} \iint\!\!\!\iint_{-\infty}^{\infty} \delta_{ijklm}(-\omega^{(4)};\omega_1,\omega_2,\omega_3,\omega_4) \nonumber \\
    &\qquad\qquad\times \tilde{F}(\omega_1)\tilde{F}(\omega_2)\tilde{F}(\omega_3)\tilde{F}(\omega_4)
                        \ee^{-\ii(\omega_1+\omega_2+\omega_3+\omega_4) t} \nonumber \\
    &\qquad\qquad\times \dd \omega_1 \dd \omega_2 \dd \omega_3 \dd \omega_4,
\end{align}
\label{freq-domain}%
\end{subequations}
where $\omega^{(n)} = \omega_1 + \omega_2 + \cdots + \omega_n$ and
\begin{equation}
    \tilde{F}(\omega) = \frac{1}{2\pi} \int_{-\infty}^\infty F(t) \ee^{\ii\omega t} \dd t.
\end{equation}
Using the notation of \citeauthor{Olsen1985},~\cite{Olsen1985} the frequency-dependent (hyper-)polarizabilities are
the linear and nonlinear response functions,
\begin{subequations}
\begin{align}
    &\alpha_{ij}(-\omega;\omega) = -\bbrakket{\hat{\mu}_i;\hat{\mu}_j}_\omega, \\
    &\beta_{ijk}(-\omega^{(2)};\omega_1,\omega_2) = \bbrakket{\hat{\mu}_i;\hat{\mu}_j,\hat{\mu}_k}_{\omega_1,\omega_2}, \\
    &\gamma_{ijkl}(-\omega^{(3)};\omega_1,\omega_2,\omega_3) = \nonumber \\
    &\qquad\quad -\bbrakket{\hat{\mu}_i;\hat{\mu}_j,\hat{\mu}_k,\hat{\mu}_l}_{\omega_1,\omega_2,\omega_3}, \\
    &\delta_{ijklm}(-\omega^{(4)};\omega_1,\omega_2,\omega_3,\omega_4) = \nonumber \\
    &\qquad\qquad\  \bbrakket{\hat{\mu}_i;\hat{\mu}_j,\hat{\mu}_k,\hat{\mu}_l,\hat{\mu}_m}_{\omega_1,\omega_2,\omega_3,\omega_4}.
\end{align}
\label{rspfun}%
\end{subequations}
The response functions of the right-hand sides can in principle be evaluated analytically with a wide range of
quantum chemical methods using response theory,~\cite{Helgaker2012} although we are not aware of any implementation
beyond cubic response---i.e., beyond the second hyperpolarizability $\gamma_{ijkl}$.

With the electric field polarized along a specific axis, say $j$, the ``diagonal'' components of the dipole
responses, $\mu_{ijj\cdots j}^{(n)}(t)$, can be extracted from $\mu_i(t)$
recorded during simulations using the central difference formulas,
\begin{subequations}
\begin{align}
    &\mu^{(1)}_{ij}(t) \approx \frac{8 \Delta^-_i(t,E_j) - \Delta^-_i(t,2E_j)} {12 E_j}, \\
    &\mu^{(2)}_{ijj}(t) \approx \frac{16 \Delta^+_i(t,E_j)  - \Delta^+_i(t,2E_j) - 30 \mu^0_i }{24 E_j^2} , \\
    &\mu^{(3)}_{ijjj}(t) \approx \frac{-13 \Delta^-_i(t,E_j) + 8\Delta^-_i(t,2E_j) - \Delta^-_i(t,3E_j)}{48 E_j^3}, \\
    &\mu^{(4)}_{ijjjj}(t) \approx  \frac{1}{144 E^4_j} \bigg{(} -39  \Delta^+_i(t,E_j) + 12  \Delta^+_i(t,2E_j) \nonumber \\ 
                         &\qquad\qquad\qquad\qquad-  \Delta^+_i(t,3E_j) +56\mu^{0} \bigg{)}.
\end{align}
\label{FFPT}%
\end{subequations}
The truncation error is $\mathcal{O}(E_j^4)$ in each case, and
\begin{equation}
\Delta^\pm_i(t,E_j) = \mu_i(t, E_j) \pm \mu_i(t, -E_j),
\end{equation}
is the sum/difference of the time-dependent dipole moments computed with opposite polarization directions and same field strength $E_j$.
One can now use different choices for $F(t)$ to obtain the frequency-dependent response functions, using either the frequency-domain
expressions ~\eqref{freq-domain} or those in the time-domain ~\eqref{time-domain2}.


\subsection{Ramped continuous wave approach} \label{sec:rcw}

As the name suggests, the RCW approach uses a continuous wave, i.e., $F(t) = \cos(\omega t)$.
This choice allows us to perform the Fourier transformations of Eq. ~\eqref{freq-domain} analytically to obtain
\begin{subequations}
\begin{align}
    &\mu^{(1)}_{ij}(t) = \alpha_{ij}(-\omega; \omega) \cos(\omega t),\label{alpha_res} \\
    &\mu^{(2)}_{ijj}(t) = \frac{1}{4} [ \beta_{ijj}^\text{SHG}(\omega) \cos(2\omega t)
                  + \beta_{ijj}^\text{OR}(\omega) ],\label{beta_res} \\
    &\mu^{(3)}_{ijjj}(t) =  \frac{1}{24} [\gamma_{ijjj}^\text{THG}(\omega) \cos(3\omega t) \nonumber \\
    &\qquad\qquad\quad + 3\gamma_{ijjj}^\text{DFWM}(\omega) \cos(\omega t) ],  \label{gamma_res}  \\
    &\mu^{(4)}_{ijjjj}(t) =  \frac{1}{192}[ \delta_{ijjjj}^\text{FHG}(\omega) \cos(4\omega t) \nonumber \\
    &\qquad\qquad\qquad\quad + 4\delta_{ijjjj}^\text{FSHG}(\omega) \cos(2\omega t) \nonumber   \\
    &\qquad\qquad\qquad\quad + 3\delta_{ijjjj}^\text{HOR}(\omega) ], \label{delta_res} 
\end{align}
\label{RCW_fit}%
\end{subequations}
where the (hyper-)polarizabilities are assumed real, a valid assumption given that $\hat{H}_0$ does not contain static magnetic fields,
and
\begin{subequations}
\begin{align}
    \beta_{ijj}^\text{SHG}(\omega) &= \beta_{ijj}(-2\omega,\omega,\omega), \\
    \beta_{ijj}^\text{OR}(\omega) &= \beta_{ijj}(0,\omega,-\omega), \\
    \gamma_{ijjj}^\text{THG}(\omega) &= \gamma_{ijjj}(-3\omega; \omega, \omega, \omega), \\
    \gamma_{ijjj}^\text{DFWM}(\omega) &= \gamma_{ijjj}(-\omega; \omega, \omega, -\omega), \\
    \delta_{ijjjj}^\text{FHG}(\omega) &= \delta_{ijjj}(-4\omega; \omega, \omega, \omega, \omega), \\
    \delta_{ijjjj}^\text{HSHG}(\omega) &= \delta_{ijjj}(-2\omega; \omega, \omega, \omega, -\omega), \\
    \delta_{ijjjj}^\text{HOR}(\omega) &= \delta_{ijjj}(0; \omega, \omega, -\omega, -\omega).
\end{align}
\label{hopol}%
\end{subequations}
The superscripts refer to the following nonlinear optical processes: Second harmonic generation (SHG), optical rectification (OR), third harmonic generation (THG),
degenerate four-wave mixing (DFWM), fourth harmonic generation (FHG), higher-order second harmonic generation (HSHG), and higher-order optical rectification
(HOR). With the left-hand sides known from simulations through Eqs.~\eqref{FFPT}, the Eqs.~\eqref{RCW_fit} yield
the frequency-dependent (hyper-)polarizabilites by curve fitting.

Frequency-dependent response theory, however, requires the field to be adiabatically switched on.~\cite{Olsen1985}
This can be achieved by a smooth modification of the continuous wave such that it is switched-on at $t\to -\infty$ and reaches full strength
at $t \to \infty$.
Of course, this
is impractical and a finite-time ramping of the field from zero to full strength must be applied in a way that minimizes nonadiabatic
effects, using only post-ramp signals to extract the dipole responses $\mu^{(n)}(t)$.
In Ref. \onlinecite{ding}, the adiabatic switching-on is simulated by a linear ramp lasting for one optical cycle, i.e.,
a ramping time $t_r = t_c$ where the cycle time is $t_c = \frac{2\pi}{\omega}$. We refer to this approach as the linear RCW (LRCW) approach
for which
\begin{equation}
F^\text{LRCW}(t) =
\left\{
\begin{array}{lcl}
    \frac{t}{t_r} \cos(\omega t)  & \qquad & 0 \leq t < t_r \\
    \cos(\omega t)                 & \qquad & t_r \leq t \leq t_\text{tot}.
\end{array}
\right.
\label{linear}
\end{equation}
Following the ramping phase, \citeauthor{ding}~\cite{ding} propagated the system for a further four optical cycles,
giving a total simulation time of $t_\text{tot} = 5t_c$. We will here investigate the effect of a longer ramping time
consisting of $n_r$ optical cycles, $t_r = n_r t_c$.
We note that the linear ramp is not continuously differentiable at $t=0$ and at $t=t_r$.

\begin{figure}
\includegraphics[scale=0.65]{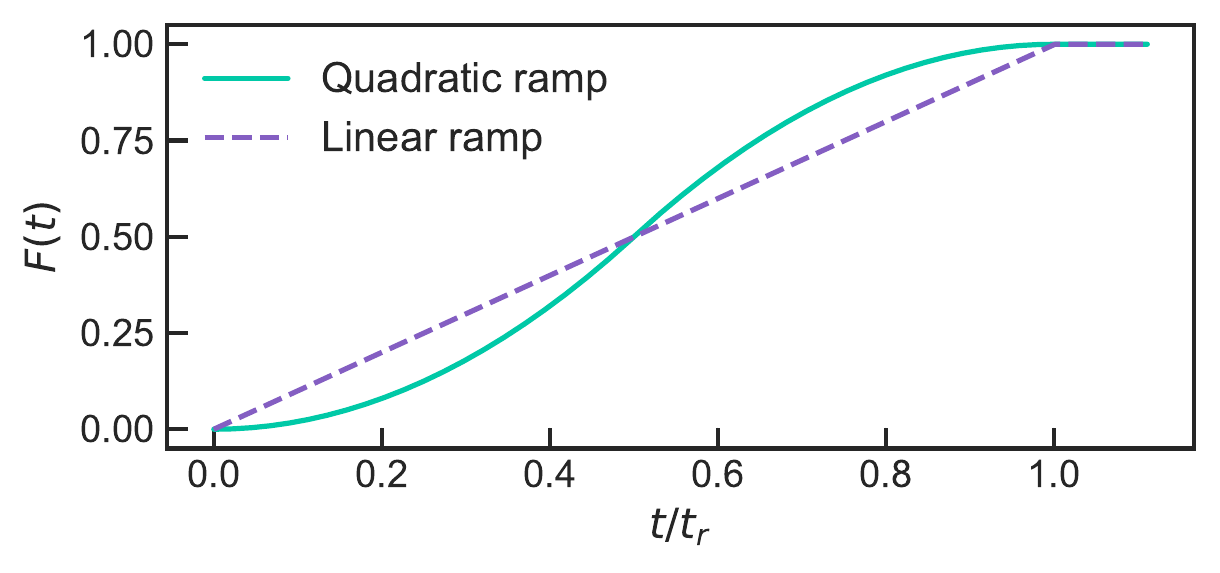}
    \caption{The time profiles $F^\text{LRCW}$ and $F^\text{QRCW}$ with $\omega=0$. \label{ramp} }
\end{figure}

In addition to extending the ramping time beyond a single optical cycle,
we investigate the \emph{quadratic ramp} to achieve a more adiabatic
switching on of the electric field. We refer to this approach as the quadratic RCW (QRCW) approach.
Specifically,
\begin{equation}
F^\text{QRCW}(t) = 
\left\{
\begin{array}{ll}
    \frac{2t^2}{t_r^2} \cos(\omega t) &  0 \leq t < \frac{t_r}{2} \\
    \frac{t_r^2 - 2(t-t_r)^2}{t_r^2} \cos(\omega t)  &   \frac{t_r}{2} \leq t < t_r \\
    \cos(\omega t)  &    t_r \leq t \leq t_\text{tot}
\end{array}
\right. \label{quadratic}
\end{equation}
which is continuously differentiable at both $t=0$ and $t=t_r$.
As illustrated in Fig.~\ref{ramp}, the quadratic ramp provides a gentler increase of the
electric field than the linear ramp for small $t$ and near $t=t_r$, albeit with a more rapid increase around $t = \frac{t_r}{2}$.
A similar sigmoid (Fermi-like) function has been used 
to obtain electronic ground states by adiabatically switching on electronic interactions,
thus providing an alternative to imaginary-time propagation.~\cite{Hermanns2012}

\subsection{Pulsed wave approach}

\citeauthor{PW}~\cite{PW} proposed an alternative approach to the extraction of linear and nonlinear properties from simulations
of electron dynamics driven by a laser pulse rather than by a continuous wave. We refer to this approach as the pulsed wave (PW)
approach, which starts from the time-domain responses, Eqs.~\eqref{time-domain2}. Retardation effects are neglected and the polarization
thus is considered local in time, i.e.,
\begin{align}
    \mu^{(n)}(t) &= \iint\!\!\cdots\!\!\int_{-\infty}^{\infty} \alpha^{(n)}(t-t_1,t-t_2,\cdots,t-t_n) \nonumber \\
    &\qquad\quad\times F(t_1)F(t_2)\cdots F(t_n) \dd t_1 \dd t_2 \cdots \dd t_n \nonumber \\
    &\approx
    \alpha^{(n)}(t) F(t)^n,
\end{align}
where the cartesian indices have been omitted for notational convenience.
Following \citeauthor{PW},~\cite{PW} the time-dependence of the finite laser pulse is described using a trigonometric envelope,
which provides a well-defined approximation to the gaussian envelope typically used in experimental work,~\cite{Barth2009}
\begin{equation}
\label{PW}
    F^\text{PW}(t) = \sin^2 \left(\frac{\pi t}{t_\text{tot}}\right) \cos(\omega t),   \qquad   0 \leq t \leq t_\text{tot}.
\end{equation}

The finite duration of the laser pulse implies that the frequency distribution is broadened around the carrier frequency $\omega$.
This, in turn, implies that $\alpha^{(n)}(t)$ contains (hyper-)polarizabilities in a range of frequencies and, therefore,
a filtering procedure must be applied to extract the proper nonlinear response functions in the frequency domain.
If the frequency distribution of the laser pulse is sufficiently sharply centered
at the carrier frequency---i.e., if the pulse duration is sufficiently long---the
linear polarizability dominates the time signal and can be found by a direct fitting of the signal to the time profile of the pulse,
\begin{equation}
    \mu^{(1)}_{ij}(t) = \alpha_{ij}(-\omega;\omega) F^\text{PW}(t). 
\end{equation}
This is essentially the same procedure used for a monochromatic continuous wave, Eq.~\eqref{alpha_res}, above.

For the hyperpolarizabilities, where more than one frequency component is present in the signal, the individual frequency components
are separated by means of a Fourier filtering procedure: $\mu^{(n)}(t)$ is Fourier transformed to the frequency domain to obtain
\begin{equation}
    \tilde{\mu}^{(n)}(\omega^\prime) = \frac{1}{2\pi} \int_{-\infty}^{\infty} \mu^{(n)}(t) \ee^{\ii\omega^\prime t} \dd t,
\end{equation}
which is subsequently transformed back to the time domain using a suitably chosen frequency window specified
as a positive integer multiple $k$ of the carrier frequency $\omega$:
\begin{align}
    \mu^{(n)}(t; k\omega) &= \int_{-(k+1)\omega}^{-(k-1)\omega}  \tilde{\mu}^{(n)}(\omega^\prime)  \ee^{-\ii\omega^\prime t}  \dd\omega^\prime
    \nonumber \\
    &+ \int_{(k-1)\omega}^{(k+1)\omega}  \tilde{\mu}^{(n)}(\omega^\prime)  \ee^{-\ii\omega^\prime t} \dd\omega^\prime .
\end{align}
The same procedure is applied to $F^\text{PW}(t)$:
\begin{align}
    F^\text{PW}(t; k\omega) &= \int_{-(k+1)\omega}^{-(k-1)\omega}  \tilde{F}^\text{PW}(\omega^\prime)  \ee^{-\ii\omega^\prime t}  \dd\omega^\prime
    \nonumber \\
    &+ \int_{(k-1)\omega}^{(k+1)\omega}  \tilde{F}^\text{PW}(\omega^\prime)  \ee^{-\ii\omega^\prime t} \dd\omega^\prime ,
\end{align}
The frequency-dependent hyperpolarizabilities are then acquired by finding the coefficient needed to fit
$\mu^{(n)}(t; k\omega)$ to $[F^\text{PW}(t; k\omega)]^n$. Thus, the first hyperpolarizabilities are
found by curve fitting according to
\begin{subequations}
\begin{align}
    &\mu^{(2)}_{ijj}(t; 0) = \frac{1}{4} \beta_{ijj}^{\text{OR}}(\omega) [F^\text{PW}(t;0)]^2, \\
    &\mu^{(2)}_{ijj}(t; 2\omega)  =  \frac{1}{4} \beta^{\text{SHG}}_{ijj}(\omega) [F^\text{PW}(t;2\omega)]^2,
\end{align}
\end{subequations}
the second hyperpolarizabilites according to 
\begin{subequations}
\begin{align}
    \mu^{(3)}_{ijjj}(t; \omega)  =  \frac{1}{8} \gamma^{\text{DFWM}}_{ijjj}(\omega)  [F^\text{PW}(t;\omega)]^3, \\
    \mu^{(3)}_{ijjj}(t; 3\omega) = \frac{1}{24} \gamma^{\text{THG}}_{ijjj}(\omega) [F^\text{PW}(t;3\omega)]^3,
\end{align}
\end{subequations}
and the third hyperpolarizabilites according to
\begin{subequations}
\begin{align}
    &\mu^{(4)}_{ijjjj}(t; 0) = \frac{1}{64} \delta^{\text{HOR}}_{ijjjj}(\omega) [F^\text{PW}(t;0)]^4, \\
    &\mu^{(4)}_{ijjjj}(t; 2\omega) = \frac{1}{48} \delta^{\text{HSHG}}_{ijjjj}(\omega) [F^\text{PW}(t;2\omega)]^4, \\
    &\mu^{(4)}_{ijjjj}(t; 4\omega) = \frac{1}{192} \delta_{ijjjj}^{\text{FHG}}(\omega)[F^\text{PW}(t;4\omega)]^4.
\end{align}
\end{subequations}

\section{Computational details}
\label{sec:compdet}

The time-dependent Schr\"{o}dinger equation \eqref{TDSE} is solved approximately using 
the time-dependent configuration-interaction singles (TDCIS)~\cite{Foresman1992,Klamroth2003} method,
the time-dependent coupled-cluster singles-and-doubles (TDCCSD)~\cite{pedersen_symplectic_2019} method,
the second-order approximate time-dependent coupled-cluster (TDCC2)~\cite{Christiansen1995,omp2} model,
and the time-dependent orbital-optimized second-order M{\o}ller-Plesset (TDOMP2)~\cite{Pathak2020,omp2} model.
For the nonvariational methods, the dipole moment is computed using the inherently real expectation-value
functional proposed in Refs.~\onlinecite{pedersen_symplectic_2019} and \onlinecite{Pedersen1997}.

\begin{figure*}
    \subfloat[\centering Linear ramp]{{\includegraphics[scale=0.60]{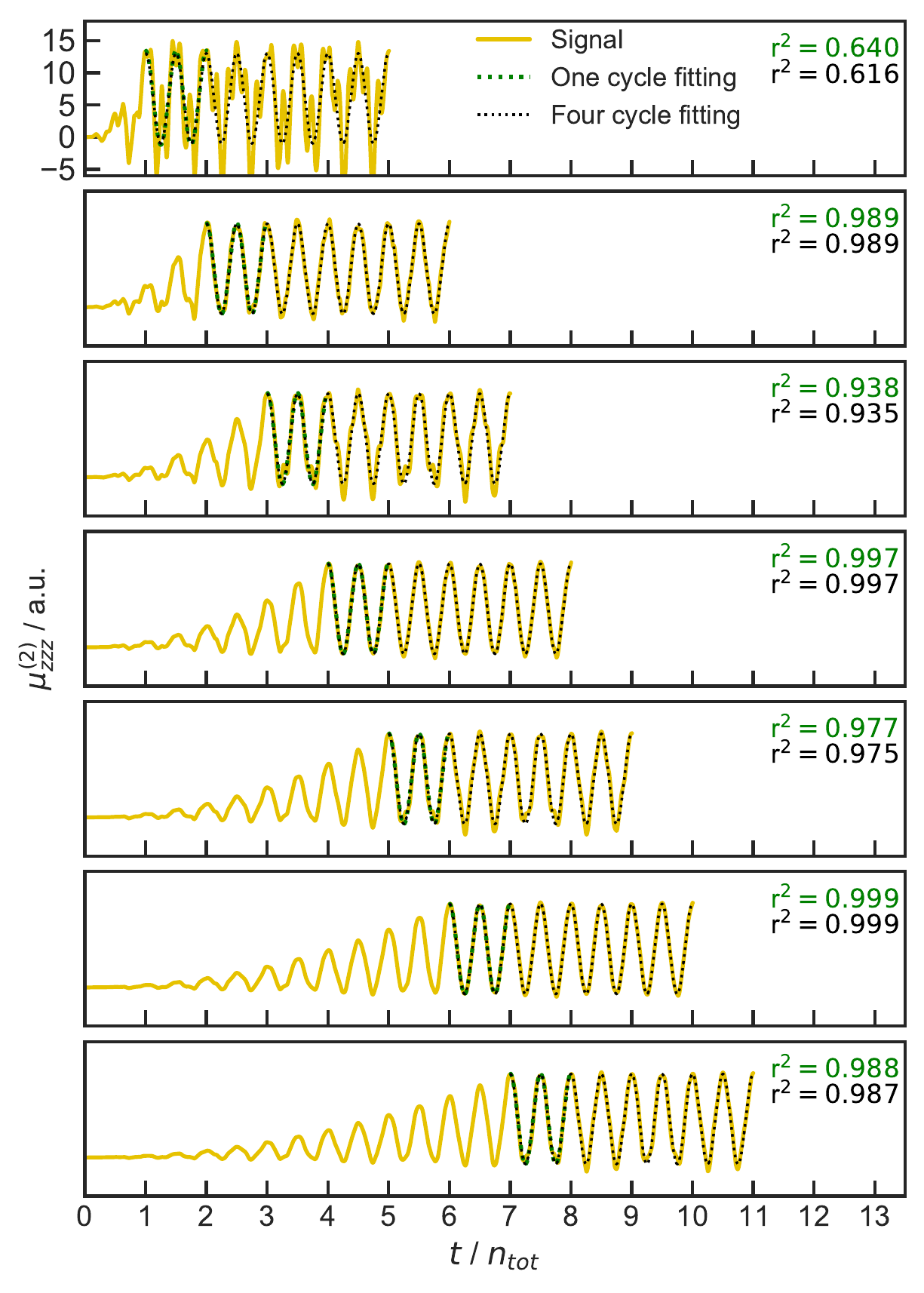} }}
    \subfloat[\centering Quadratic ramp]{{\includegraphics[scale=0.60]{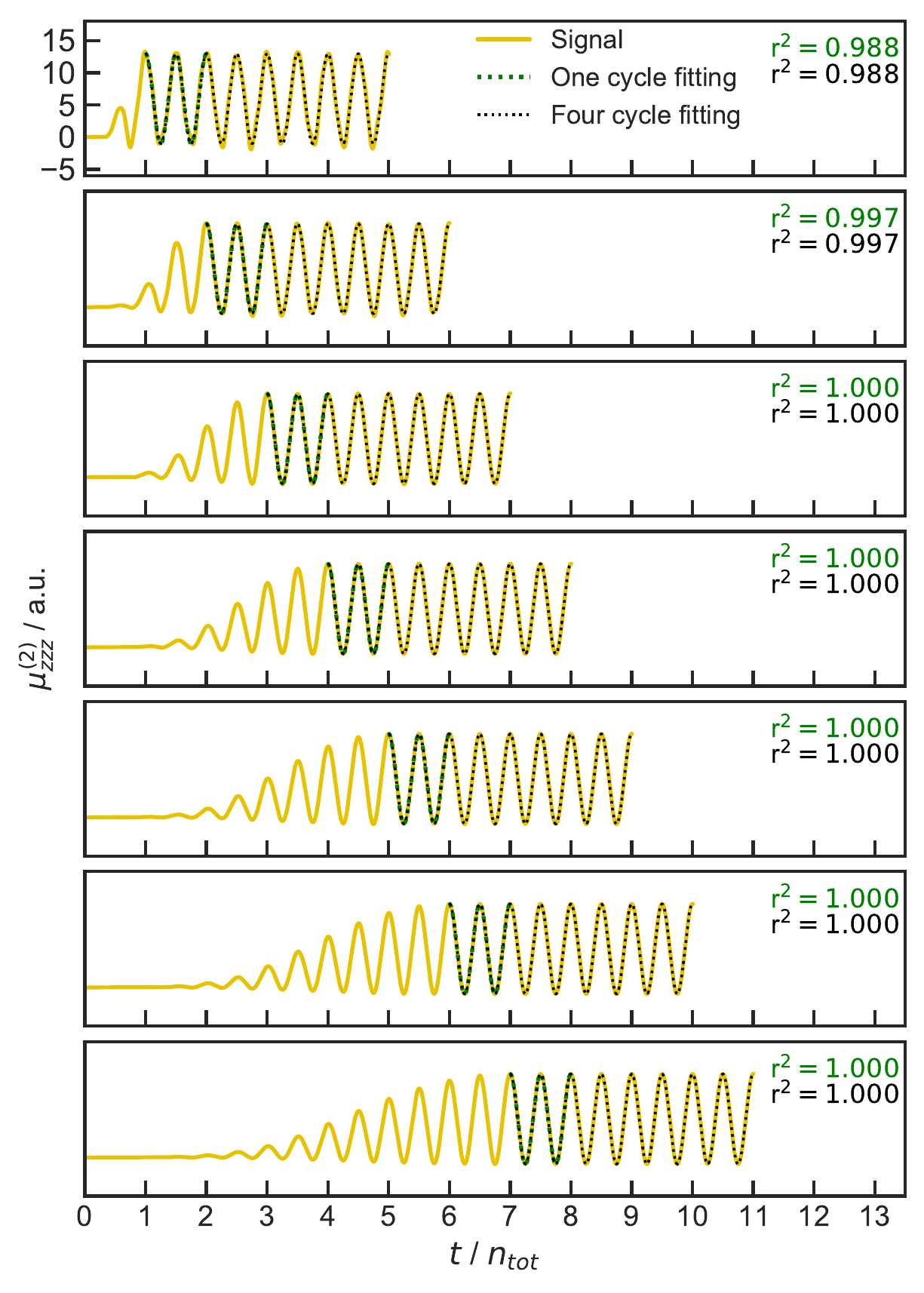} }}
    \caption{The second-order dipole response obtained from TDCCSD simulations using, from top to bottom, $n_r =(1, 2, 3, 4 ,5 ,6 ,7)$ linear ramp (a) or
    quadratic ramp (b) cycles followed by four post-ramp cycles of propagation time for NH$_3$. The fitting in black is done on all four post-ramp cycles, the fitting in green is done on one post-ramp cycle. All plots are on the same scale.\label{hyppol}}
\end{figure*}

\begin{figure*}
\begin{center}
\begin{subfigure}{0.44\textwidth}
\includegraphics[width=\textwidth]{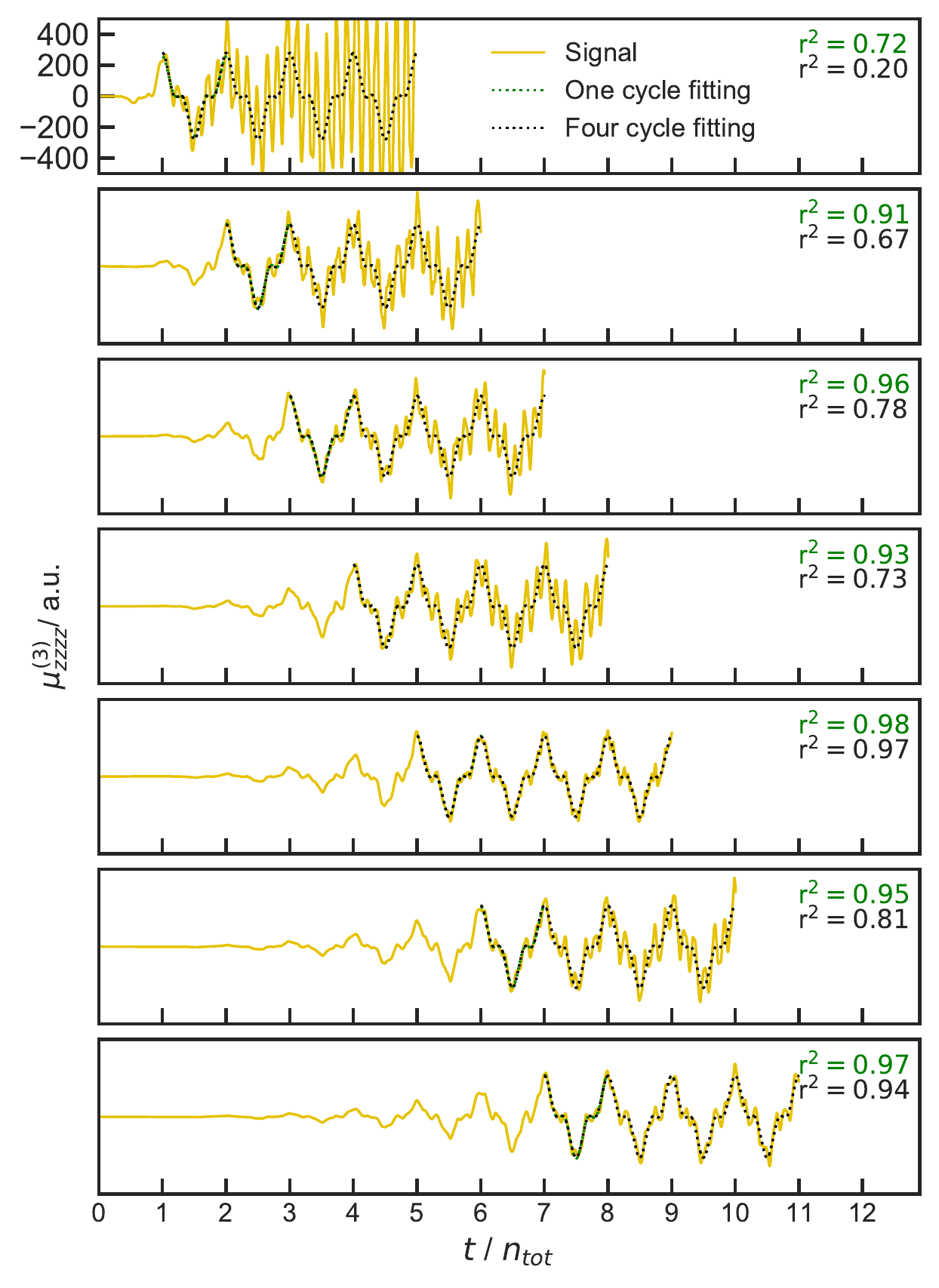} 
\caption{\centering H$_2$O:  Linear ramp}
\label{sec_hyp_a}
\end{subfigure}
%
\begin{subfigure}{0.44\textwidth}
\includegraphics[width=\textwidth]{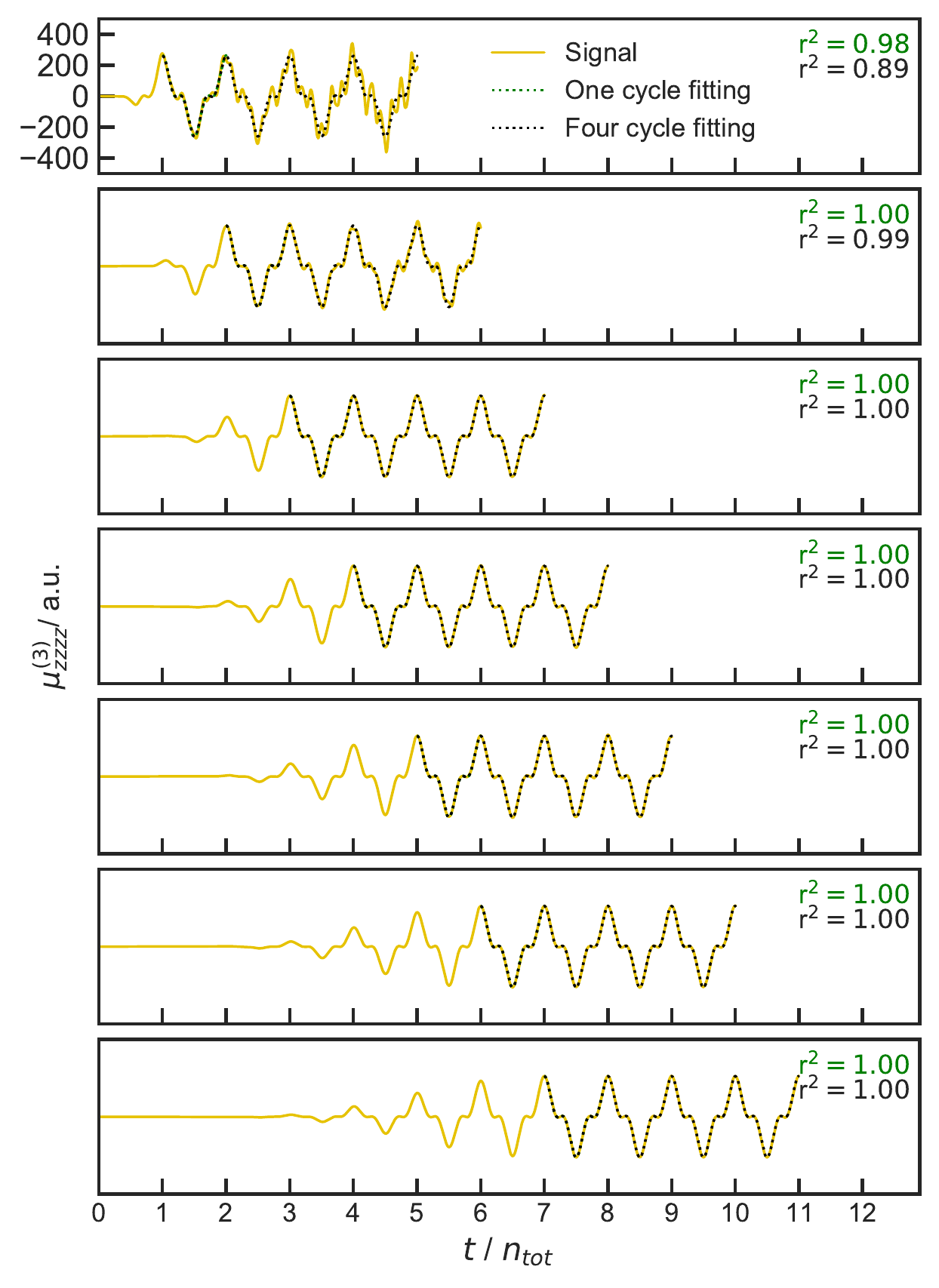} \\
\caption{\centering H$_2$O:  Quadratic ramp}
\label{sec_hyp_b}
\end{subfigure}
\hfill
\begin{subfigure}{0.44\textwidth}
\includegraphics[width=\textwidth]{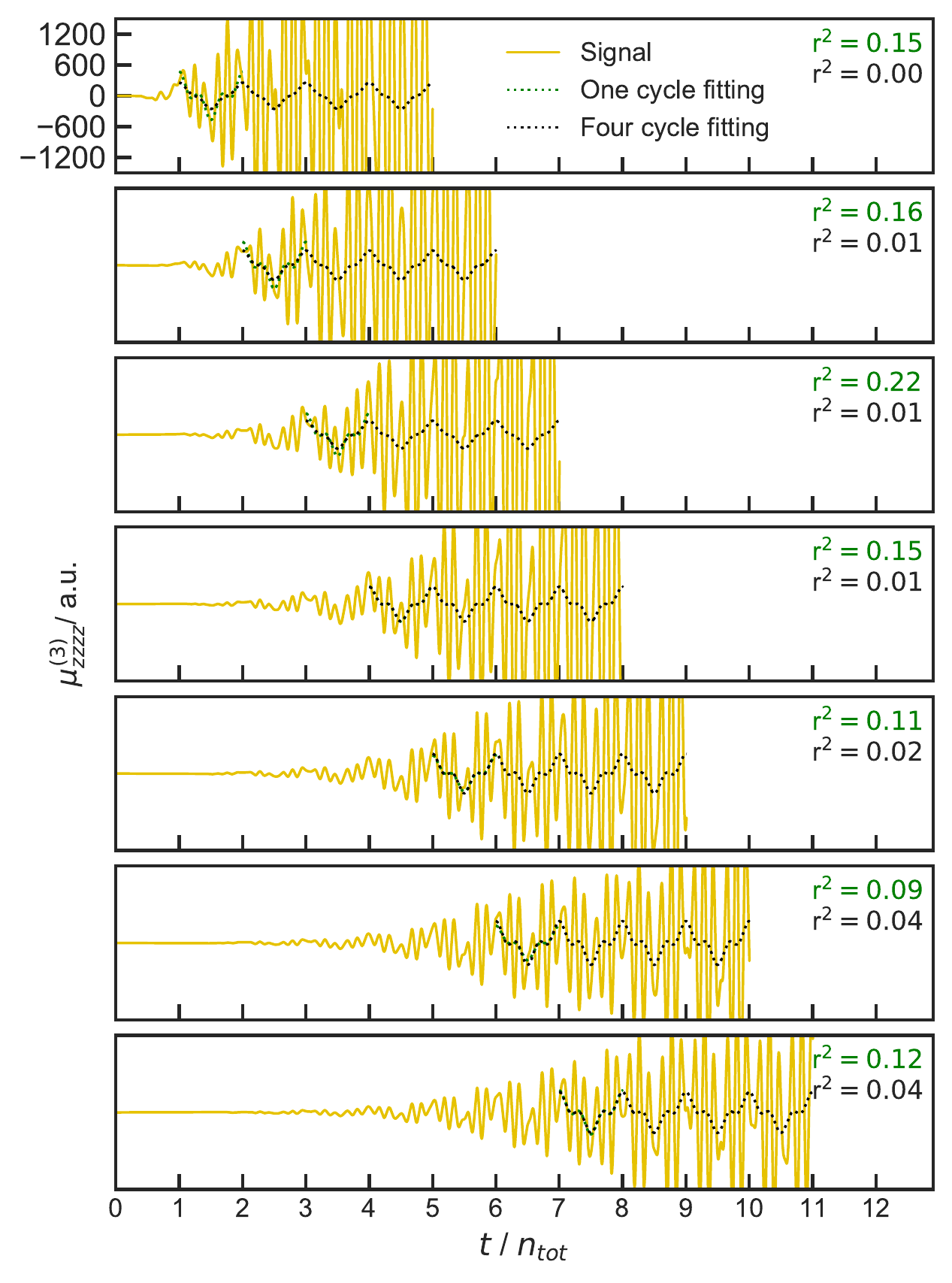} 
\caption{\centering CH$_4$: Linear ramp}
\label{sec_hyp_c}
\end{subfigure}
\begin{subfigure}{0.44\textwidth}
\includegraphics[width=\textwidth]{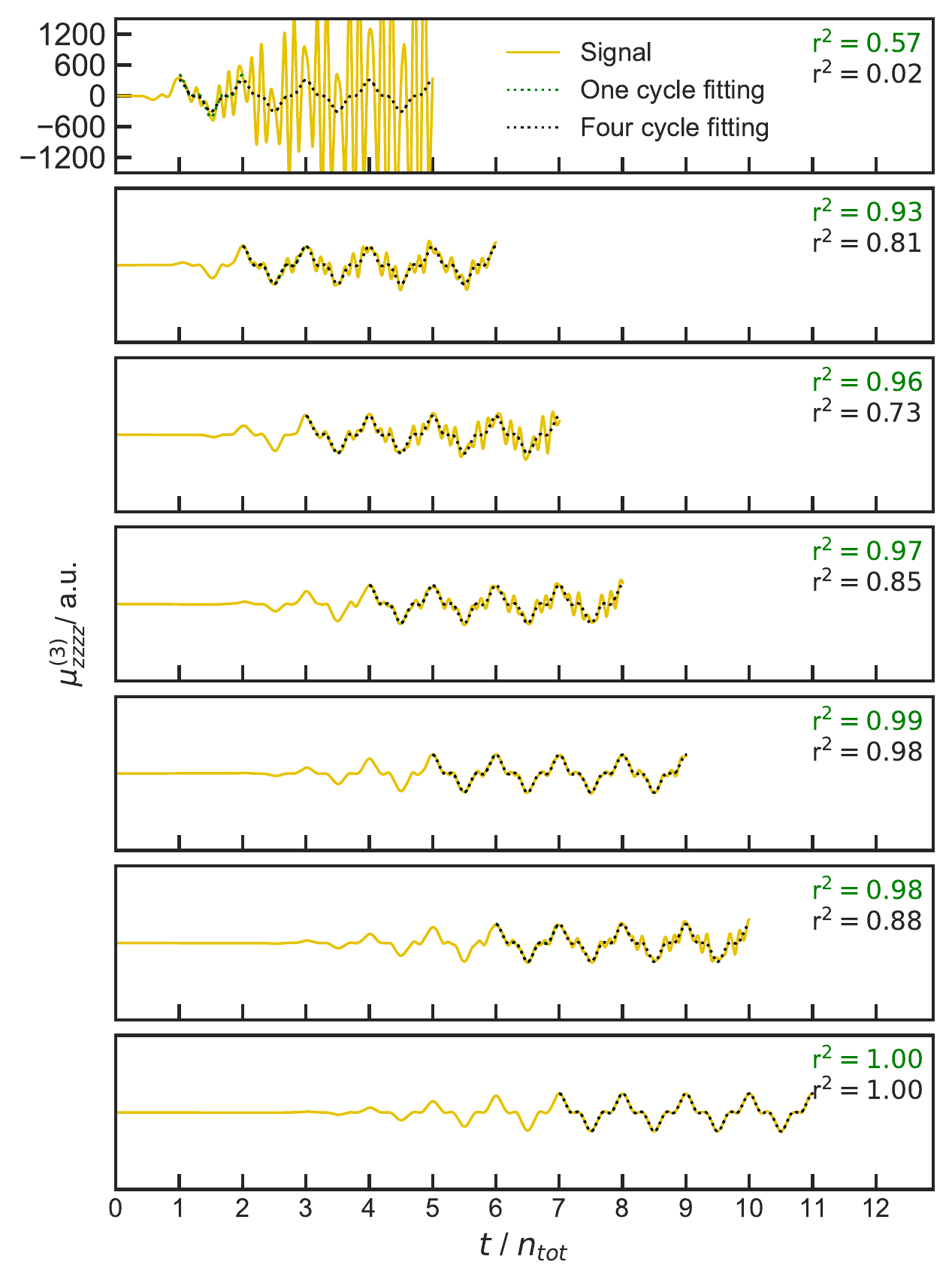} 
\caption{\centering CH$_4$: Quadratic ramp}
\label{sec_hyp_d}
\end{subfigure}
\caption{The third-order dipole response obtained from TDCCSD simulations using, from top to bottom, $n_r =(1, 2, 3, 4 ,5 ,6 ,7)$ linear ramp (a) or quadratic ramp (b) cycles followed by four post-ramp cycles of propagation for H$_2$O and Ne. The fitting in black is done on all four post-ramp cycles, the fitting in green is done on one post-ramp cycle.  All four plots are on the same scale for each system.\label{sec_hyppol}}
\end{center}
\end{figure*}

We test the RCW and PW approaches using the same ten-electron systems as in Ref.~\onlinecite{omp2}, namely
Ne, HF, H$_2$O,  NH$_3$, and CH$_4$. The geometries of these molecules can be found in the supplementary material.
The d-aug-cc-pVDZ~\cite{woon_gaussian_1994} basis set is used for Ne, while the aug-cc-pVDZ~\cite{kendall_electron_1992} basis set 
is used for the four remaining systems. The basis set definitions are taken from the Basis Set Exchange.\cite{bse}
The carrier frequencies are chosen in accord with Ref.~\onlinecite{omp2}:
Ne: $\omega = 0.1$, HF: $\omega = 0.1$,  H$_2$O: $\omega = 0.0428$,  NH$_3$:  $\omega = 0.0428$,  and CH$_4$: $\omega = 0.0656\,\text{a.u.}$
These frequencies come in at less than one third of the first dipole-allowed excitation energy for each system, enabling properties up to (at least) the third hyperpolarizability to be reliably extracted.

The Hartree-Fock reference orbitals and Hamiltonian integrals are calculated using the Python-based Simulations of Chemistry Framework\cite{pyscf} (PySCF) with the gradient norm convergence threshold set to $10^{-10}\,\text{a.u.}$

The ground states are computed using a locally developed closed-shell spin-restricted code \cite{HyQD}, all computed with a residual norm convergence criteria
of $10^{-12}\,\text{a.u.}$ 
The equations of motion are integrated using the sixth order (three-stage, $s=3$) symplectic Gauss-Legendre
integrator~\cite{integrator} as described in Ref.~\onlinecite{pedersen_symplectic_2019} with
a time step of $\Delta t = 0.01\,\text{a.u.}$ and the residual norm convergence criterion set to $10^{-10}\,\text{a.u.}$ for the implicit equations.
The least-squares curve fitting, using the Levenberg-Marquardt algorithm, as implemented in the \emph{optimize} module of
SciPy~\cite{2020SciPy-NMeth} is used to extract the (hyper-)polarizabilities.

The coupled-cluster response data are computed using the
Dalton quantum chemistry package~\cite{dalton1,dalton2,Koch1996,Halkier1997,Christiansen1998,Hattig1997,Hattig1998}
with the following gradient/residual norm convergence criteria: $10^{-10}\,\text{a.u.}$ for the Hartree-Fock reference orbitals, $10^{-10}\,\text{a.u.}$
for the CC ground-state residual norms, and $10^{-8}\,\text{a.u.}$ for the response equations. The CIS response data are
computed using sum-over-states expressions.~\cite{Olsen1985,Orr1971}

For all simulations, the electric-field strengths $E = \pm 0.001, \pm  0.002,  \pm  0.003\,\text{a.u.}$ are used in order to reduce numerical noise
for higher-order properties. The electric-field strengths should be of a magnitude where \emph{both} the errors associated with numerical noise \emph{and}
the errors arising from numerical truncation and nonadiabatic effects remain small. 
\citeauthor{ding}~\cite{ding} explored field strengths in the range $0.0005\,\text{a.u.}$ to $0.005\,\text{a.u.}$ and found $E = 0.002\,\text{a.u.}$
to provide the most accurate results. \citeauthor{PW}~\cite{PW} used field strengths from $E \approx 0.0002\,\text{a.u.}$ to
$E \approx 0.002\,\text{a.u.}$, and did not find the PW approach to be sensitive within this range. 


\section{Results}
\label{sec:results}

\subsection{Time evolution of the nonlinear dipole responses in the RCW approach} \label{time-evolution}


The accuracy of the (hyper-)polarizabilities obtained using the RCW approach depends on how closely the time-domain dipole responses $\mu^{(n)}(t)$ \emph{actually} are to their expected forms, expressed by Eqs.~\eqref{alpha_res} -- \eqref{delta_res}. Therefore, we start by comparing the signals extracted after linear and quadratic ramping. The motivation for this is twofold: Firstly, we wish to investigate if the deviations previously observed~\cite{ding} for time signals of nonlinear properties can be alleviated with the closer-to-adiabatic quadratic ramp. Secondly, we wish to investigate the effect of varying the ramping time $t_r = n_r t_c$ and the propagation time $t_p = n_p t_c$. Their relative importance will be assessed, and the most favorable ratio between $n_r$ and $n_p$ will be determined. When the time parameters are explicitly specified, the approach will be denoted RCW$(n_r, n_p)$. Results obtained with the TDCCSD method will be presented and discussed; analogous results with the TDCC2, TDOMP2, and TDCIS methods are given in the supplementary material.


The systems are ramped using either the linear ramp profile $F^\text{LRCW}$ [Eq.~\eqref{linear}] or the quadratic ramp profile $F^\text{QRCW}$ [Eq.~\eqref{quadratic}].  The ramp duration is increased in increments of one optical cycle, up to a maximum of seven optical cycles ($n_r = 1, 2, \cdots ,7$), followed by four optical cycles of propagation ($n_p = 4$). The second- and third-order time-dependent dipole responses $\mu^{(n)}(t)$, $n=2,3$, are collected, and fitted to the expected shapes in Eqs.~\eqref{beta_res} and \eqref{gamma_res}.


The time-domain dipole response $\mu^{(2)}_{zzz}(t)$ for the NH$_3$ molecule is displayed in Fig.~\ref{hyppol}. The upper-most panel exhibits the one-cycle ramp, and for each descending panel the number of ramping cycles is increased by one. The function obtained by fitting the analytic form over the range of four cycles post-ramp is displayed in black along with its coefficient of determination, $r^2$. The function obtained by fitting to one post-ramp cycle is plotted in green.

Ramping with the linear profile, we observe the high-frequency oscillations previously reported in Refs.~\citen{ding} and \citen{omp2}. The correspondence between the signal and expected form improves significantly as the linear ramping time is increased to two and three optical cycles. Increasing the linear ramp time above three optical cycles only delivers marginal improvements. The simulations conducted using the quadratic ramp, in contrast, appear to behave correctly already at the one-cycle ramp stage. Evidently, the signal can be improved either by switching to the quadratic ramp or by increasing the duration of the linear ramping. The curves fitted using one post-ramp cycle typically have $r^2$ values slightly above those obtained by fitting to four post-ramp cycles.


The second hyperpolarizability, a much smaller contribution to the total induced dipole moment, is highly sensitive to errors in the time signal. The third-order response signal, $\mu^{(3)}_{zzz}(t)$, features both the high-frequency oscillations observed for the first hyperpolarizabily and the increase of amplitude as
time progresses.
The molecule least sensitive to these effects is H$_2$O, for which results are presented in Fig.~\ref{sec_hyp_a}--\ref{sec_hyp_b}. The molecule most sensitive is CH$_4$, for which results are presented in Fig.~\ref{sec_hyp_c}--\ref{sec_hyp_d}.
The linear one-cycle ramp is clearly inadequate for describing the second hyperpolarizability for both molecules, indicating that even a small amount of nonadiabatic error dramatically reduces the correspondence of the signal with its expected form. Increasing the ramping time to two or three optical cycles greatly improves the signal, but analogously to what was observed for the first hyperpolarizability, increasing beyond three cycles does not improve the signal. Even with increased ramping time, the linear ramp does not provide an accurate description of the third order response signal, yielding at best $r^2=0.22$ for CH$_4$. The quadratic ramp profile fairs notably better, although the high-frequency oscillations and the drift of the amplitude is observed for the CH$_4$ molecule when a one-cycle ramp is employed---i.e., in contrast to what was observed for the first hyperpolarizability, a one-cycle quadratic ramp appears to be insufficient. Increasing the number of ramping cycles quickly leads to convergence, giving $r^2 = 0.99998$ when seven cycles are used for the CH$_4$ molecule. 

The gradual ramping of the electric-field strength is found to reduce the signal errors indicating
that nonadiabatic effects are the main source of error, not higher-order response contributions.
Furthermore, it is clear that the quadratic ramp aids in reaching an adiabatic description more rapidly than the linear ramp.
It is found that fitting the expected function to the signal after only one post-ramp cycle is warranted when
the simulation has ramped in a sufficiently adiabatic manner. Based on these observations,
we shall compare the RCW method to the PW method using one post-ramp cycle.


\begin{figure}
{{\includegraphics[scale=0.55]{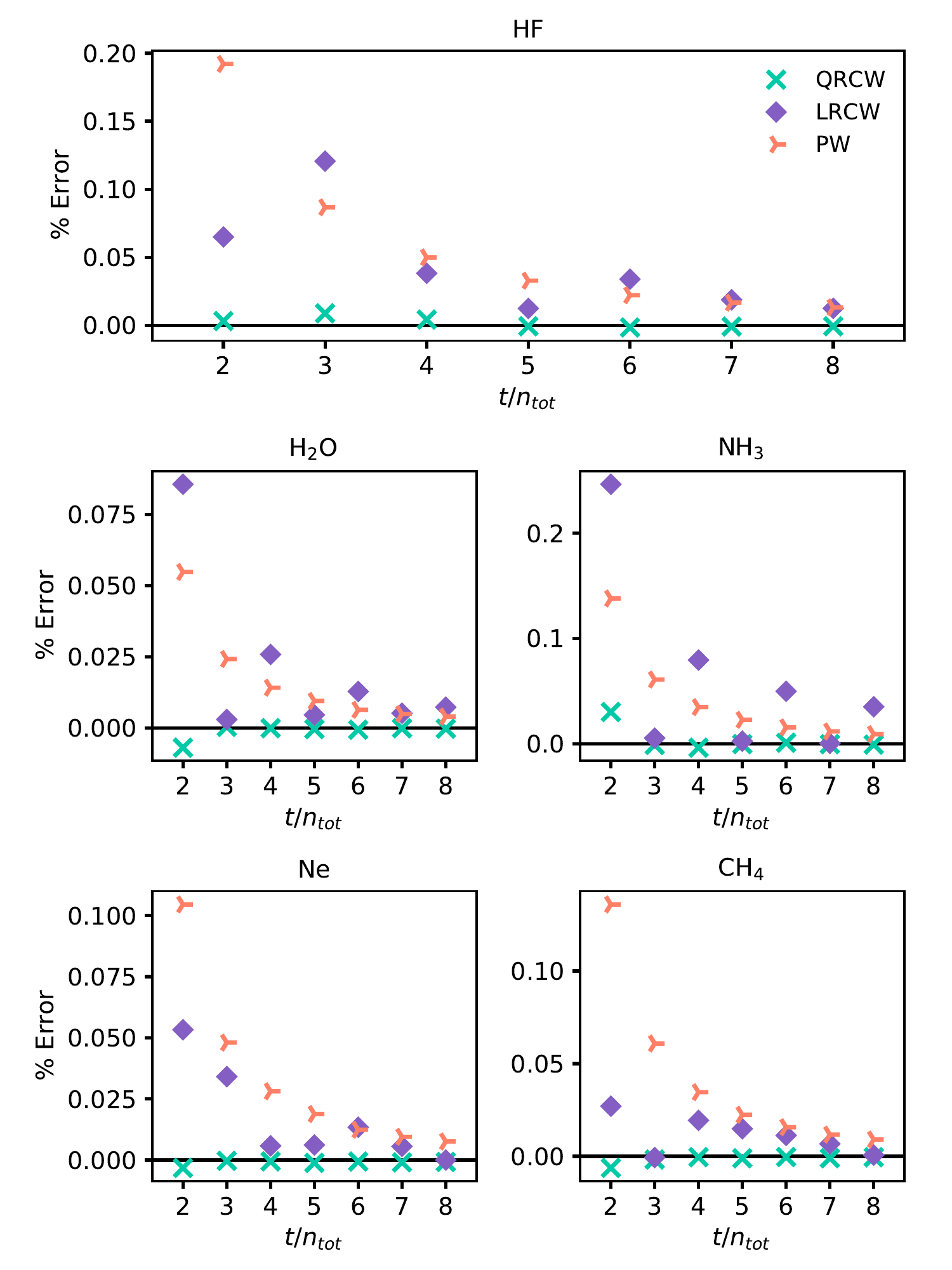} }}
\caption{Convergence towards CCSD linear response results of
    polarizabilities, $\alpha_{zz}$, 
    extracted from TDCCSD simulations using the LRCW($n_\text{tot}-1$,$1$), QRCW($n_\text{tot}-1$,$1$), and PW($n_\text{tot}$) methods.
    \label{pol_ccsd}}
\end{figure}

\subsection{Polarizability}

In the interest of gauging the accuracy of the different approaches, polarizabilities extracted using the LRCW, QRCW, and PW approach are compared to response theory calculations. By assessing the closeness of the real-time approaches to response theory at different total simulation times,
$t_{\text{tot}} = (n_r + n_p) t_c = n_\text{tot}t_c$, we may compare the accuracy achieved by the three approaches at similar computational costs. 

The RCW simulations are performed using $n_\text{tot} = n_r + 1$, i.e., using the LRCW$(n_r,1)$ and QRCW$(n_r,1)$ methods,
in line with the discussion in section \ref{time-evolution}. The first-order
dipole response is separated from the time signal using the finite difference formula \eqref{FFPT} for all approaches. 
By point-group symmetry, the polarizability tensors of Ne and CH$_4$ are equal for all Cartesian directions, and only the $zz$ component is computed. For the NH$_3$ and HF molecules, the $xx$ and $yy$ polarizability components are the same by symmetry, whereas polarizability tensors for all three Cartesian directions are computed for H$_2$O.  The polarizabilities at the CCSD level for the five systems in the $zz$ direction are displayed in Fig.~\ref{pol_ccsd}.
The other unique polarizability components at the CCSD level can be found in the supplementary material along with CC2 results. 

The QRCW approach consistently produces polarizabilities with the highest accuracy, achieving a $0.03\%$ accuracy after the minimum two cycles of total simulation time, $n_r =1,  n_p = 1$. The PW approach requires longer computational time in order to attain polarizabilities of the same accuracy as QRCW, yet it converges consistently towards the correct value with increasing cycles of simulation time. Albeit with irregular convergence behavior, the LRCW method achieves accuracies comparable to the PW approach as illustrated for the NH$_3$ and H$_2$O molecules in Fig.~\ref{pol_ccsd}. 

The errors for the extracted polarizabilities are overall small for all three approaches regardless of ramping. For the shortest ramp length and poorest performing extraction approach, the polarizability is still correct to $0.25\%$. This modest error can be reduced to below $0.0009$\% using the QRCW(7,1) method 
for all molecules.



\subsection{First hyperpolarizability}

The second-order dipole response signal is sensitive to nonadiabatic effects, as seen in Fig.~\ref{hyppol}. Adiabatic ramping, therefore,
is expected to significantly improve accuracy.

\begin{figure}
{{\includegraphics[scale=0.55]{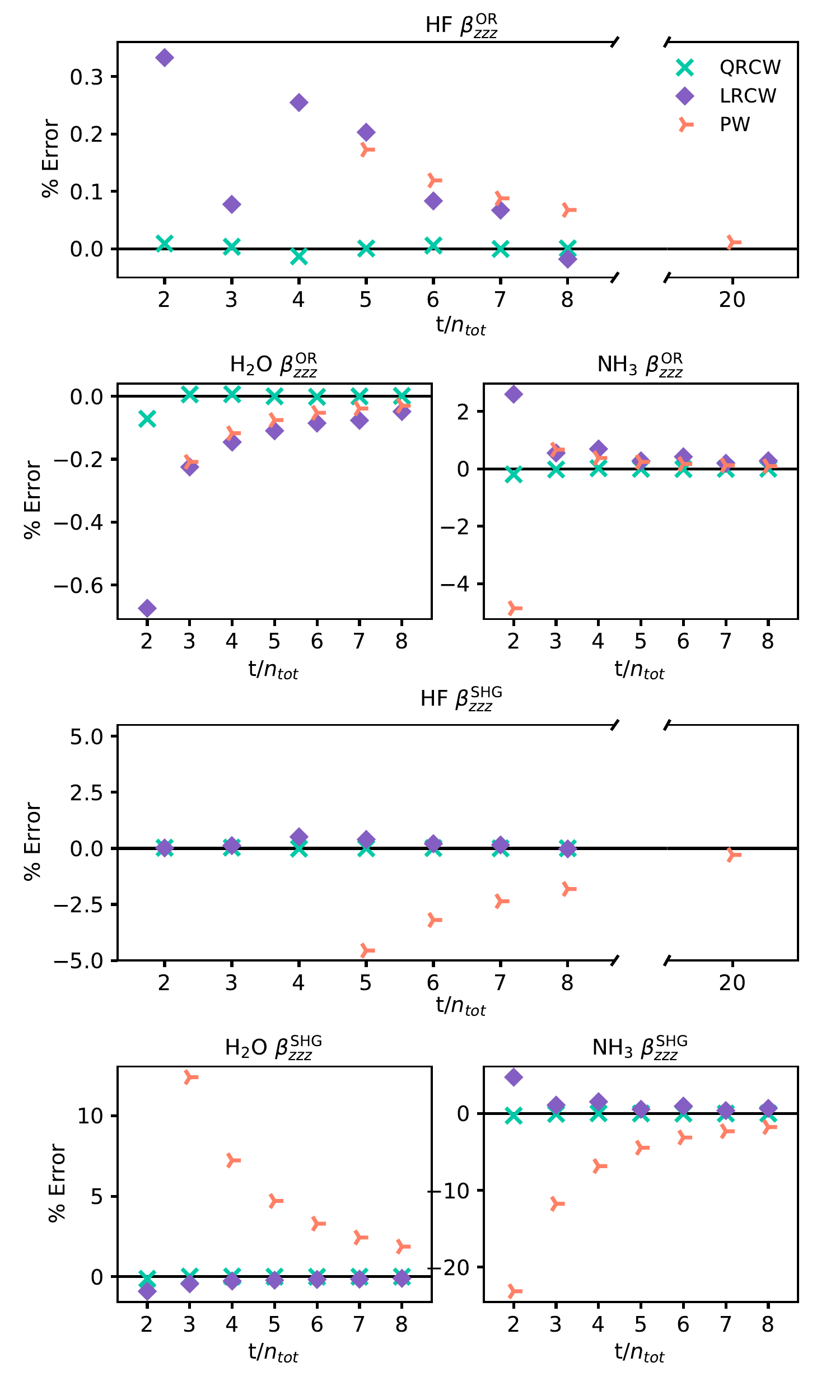} }}
\caption{Convergence towards CCSD quadratic response results of
    first hyperpolarizabilities extracted from TDCCSD simulations
    using the LRCW($n_\text{tot}-1$,$1$), QRCW($n_\text{tot}-1$,$1$), and PW($n_\text{tot}$) methods.
    \label{first_hyper} }
\end{figure}

\begin{table}
\centering
    \caption{First hyperpolarizabilities, $\beta^\text{SHG}_{zzz}$, extracted from TDCCSD simulations using the LRCW($n_r$,$n_p$) and QRCW($n_r$,$n_p$)
    methods compared with CCSD linear response results.\label{hyppol_tab}}
\begin{tabular}{l r r r }
\hline
\hline
\vspace{-0.2cm} \\
          & HF  & H$_2$O & NH$_3$ \\
\hline
LRCW(1,4) \quad & \quad 14.428  &\quad -9.683  & \quad 29.330         \\
\vspace{0.15cm}
QRCW(1,4)  & 14.372  & -9.588  & 28.022         \\
QRCW(1,1)  & 14.375  & -9.603  & 27.941         \\
QRCW(2,1)  & 14.368  & -9.590  & 28.005         \\
QRCW(3,1)  & 14.371  & -9.590  & 28.027         \\
QRCW(4,1)  & 14.372  & -9.591  & 28.020         \\
QRCW(5,1)  & 14.370   & -9.591  & 28.020       \\  
\hline 
Response   & 14.370 & -9.591  & 28.020        \\
\hline\hline
\end{tabular}
\end{table}

\begin{figure*}
\begin{center}
\begin{subfigure}{0.44\textwidth}
\includegraphics[width=\textwidth]{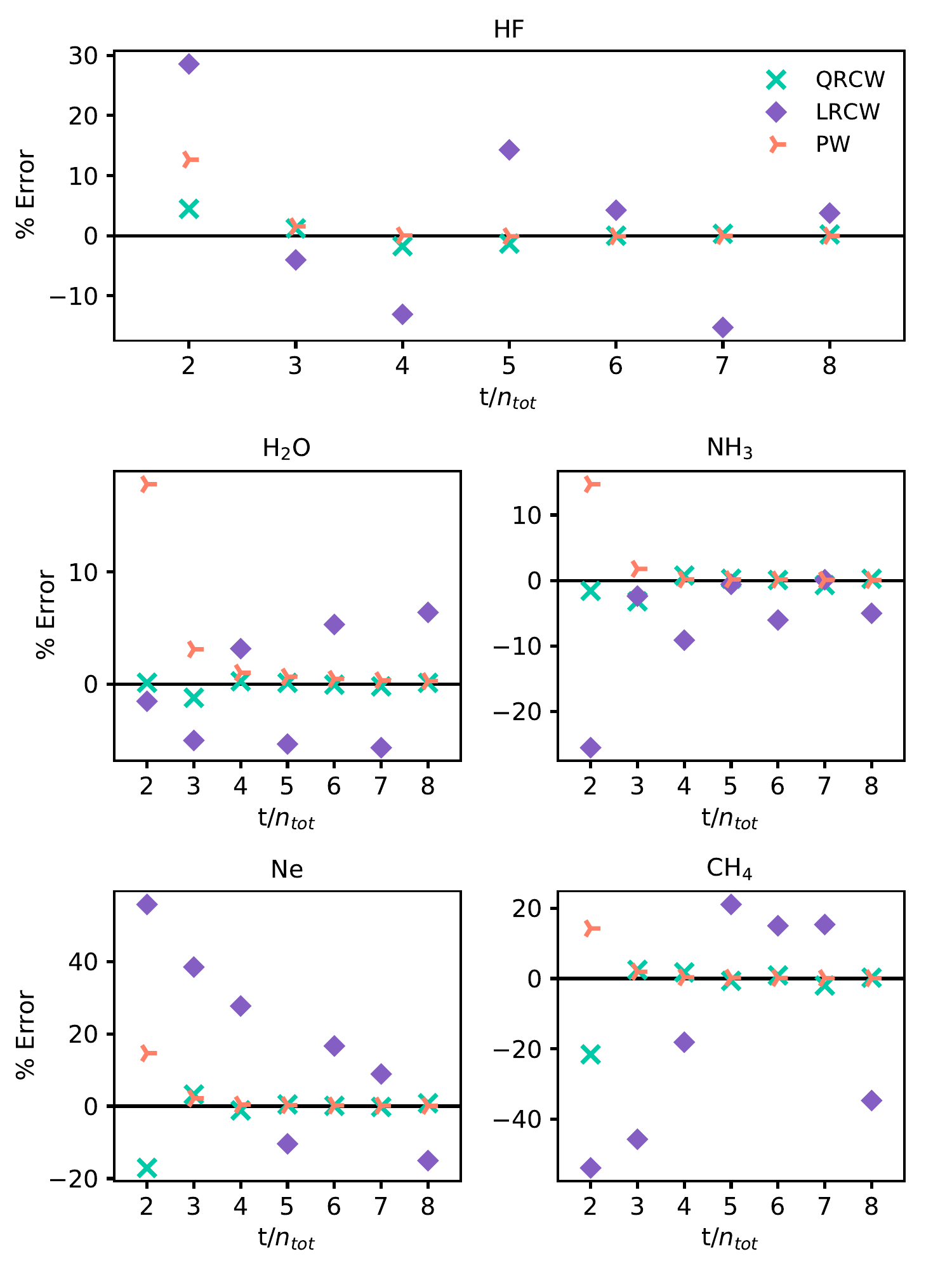} 
\caption{\centering $\gamma^\text{THG}_{zzzz}$}
\label{sec_hyp_thg}
\end{subfigure}
%
\begin{subfigure}{0.44\textwidth}
\includegraphics[width=\textwidth]{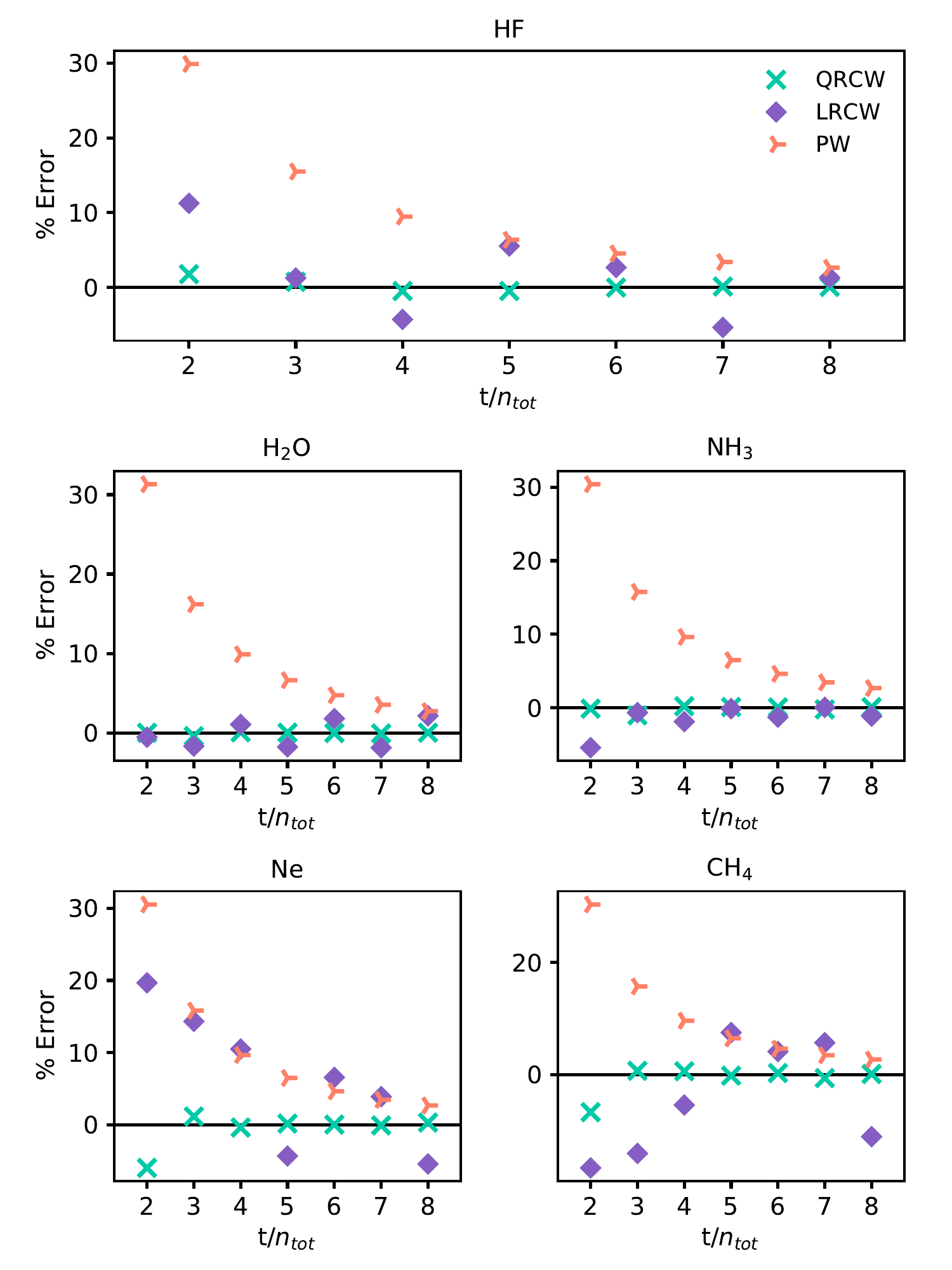} 
\caption{\centering $\gamma^\text{DFWM}_{zzzz}$}
\label{sec_hyp_dfwm}
\end{subfigure}
\caption{Convergence towards CCSD cubic response results of
    second hyperpolarizabilities extracted from TDCCSD simulations
    using the LRCW($n_\text{tot}-1$,$1$), QRCW($n_\text{tot}-1$,$1$), and PW($n_\text{tot}$) methods.
    \label{second_hyp}}
\end{center}
\end{figure*}

Due to symmetry, all the diagonal components of the first hyperpolarizability are zero for the Ne and CH$_4$ molecules. The H$_2$O and the HF molecule exhibit diagonal first hyperpolarizabilities in the $zzz$-direction, and the NH$_3$ molecule has non-vanishing hyperpolarizabilities in the $yyy$- and $zzz$-direction. The $\beta^{\text{SHG}}_{zzz}$ and $\beta^{\text{OR}}_{zzz}$ components extracted from TDCCSD simulations are
displayed in Fig.~\ref{first_hyper}. 
The $\beta^{\text{SHG}}_{yyy}$ and $\beta^{\text{OR}}_{yyy}$ components of NH$_3$ can be found in the supplementary material
along with all non-zero diagonal first hyperpolarizabilitaties at the CC2 level. 

The errors of the extracted first hyperpolarizabilities are found to be 
roughly an order of magnitude greater than for polarizabilities with $n_\text{tot}=2$.
Using the LRCW(1,1) method to extract the $\beta^{\text{SHG}}_{zzz}$ component yields the following \% errors:
HF 0.33\%, H$_2$O 0.33\%, and NH$_3$ 2.6\%. 
Using the QRCW(1,1) method reduces these to: HF 0.01\%, H$_2$O 0.07\%, NH$_3$ 0.2\%. 
We obtain hyperpolarizabilites with the smallest relative errors when applying QRCW(7,1) with accuracies of: HF 0.0006\%, H$_2$O 0.001\%, NH$_3$ 0.003\% for the $\beta^{\text{SHG}}_{zzz}$ component and: HF 0.003\%, H$_2$O 0.002\%, NH$_3$ 0.005\% for the $\beta^{\text{OR}}_{zzz}$ component. With this error reduction the first hyperpolarizabiliy tensors have accuracies comparable to the extracted polarizability tensors. 

The PW approach performs surprisingly poorly with regards to extracting the first hyperpolarizabilies as seen in Fig.~\ref{first_hyper}.
For the shortest simulation times, some of the errors obtained for the $\beta^{\text{SHG}}_{zzz}$ component are beyond the scale of the plot,
especially for the HF molecule.
To check convergence,
the total propagation time for the HF molecule was increased to PW(20), resulting in a reduction of the error from $0.07\%$ to $0.01\%$ for the $\beta^{\text{OR}}_{zzz}$ component and from $-1.8\%$ to $0.29\%$ for the $\beta^{\text{SHG}}_{zzz}$ component.
Although improved, the PW(20) results are still of significantly lower accuracy than the RCW(7,1) results, for which the errors are $0.0006\%$ for the $\beta^{\text{OR}}_{zzz}$ component and $0.004\%$ for the $\beta^{\text{SHG}}_{zzz}$ component. 


The accuracies achieved using the QRCW$(n_r, 1)$ with $n_r =(1, 2, 3 ,4, 5 )$ are compared to LRCW(1,4) and QRCW(1,4) in Table \ref{hyppol_tab}.
Depending on the demands placed on computational cost and accuracy, it is worth noting that propagating for only two optical cycles ($n_r=n_p=1$) may be sufficient: 
The QRCW(1,1) approach produces higher accuracy than the LRCW(1,4) approach.

\begin{table*}
\centering
\caption{The third harmonic generation components $\gamma^{\text{THG}}_{jjjj}$ of the second hyperpolarizabilities of Ne, HF, H$_2$O, NH$_3$ and CH$_4$ extracted from TDCCSD, TDOMP2, TDCC2, and TDCIS simulations
    compared with results from cubic response theory.
    \label{sec_pol_tab}}
\begin{tabular}{l l r r r r r r r r r r }
\hline 
\hline
\vspace{-0.1cm} \\
&  & TDCCSD & TDOMP2 & TDCC2 & TDCIS & &  TDCCSD & TDOMP2 & TDCC2 & TDCIS \\ 
\hline
\vspace{-0.2cm}\\
HF  & & \multicolumn{4}{c}{$\gamma^{\text{THG}}_{xxxx}$}  & & \multicolumn{4}{c}{$\gamma^{\text{THG}}_{zzzz}$} \vspace{0.05cm} \\ 
 \cline{3-6}  \cline{8-11} \vspace{-0.2cm} \\
& PW(3) & $-718$  & $-912$  & $-1180$  & $-236$   &  & $-509$  & $-531$  & $-685$  & $-329$   &  \\ 
& LRCW(2,1) & $-834$  & $-997$  & $-1050$  & $-196$   &  & $-539$  & $-426$  & $-588$  & $-321$   &  \\ 
& QRCW(2,1) & $-570$  & $-700$  & $-1010$  & $-227$   &  & $-511$  & $-524$  & $-643$  & $-332$   &  \\ 
 & Response  & $-625$  & $ $  & $-949$  & $-230$   &  & $-517$  & $ $  & $-653$  & $-332$   & \vspace{0.2cm} \\ 
NH$_3$  & & \multicolumn{4}{c}{$\gamma^{\text{THG}}_{yyyy}$}  & & \multicolumn{4}{c}{$\gamma^{\text{THG}}_{zzzz}$}  \vspace{0.05cm} \\ 
 \cline{3-6}  \cline{8-11}  \vspace{-0.2cm} \\
& PW(3) & $-1180$  & $-1260$  & $-1380$  & $10.3$   &  & $-7710$  & $-8930$  & $-9770$  & $-4030$   &  \\ 
& LRCW(2,1) & $-770$  & $-1850$  & $-2010$  & $-470$   &  & $-8030$  & $-9500$  & $-11000$  & $-4310$   &  \\ 
& QRCW(2,1) & $-1260$  & $-1300$  & $-1450$  & $-2.40$   &  & $-8100$  & $-9070$  & $-9990$  & $-4110$   &  \\ 
 & Response  & $-1220$  & $ $  & $-1420$  & $13.8$   &  & $-7850$  & $ $  & $-9950$  & $-4130$   &  \vspace{0.2cm} \\ 
H$_2$O  & & \multicolumn{4}{c}{$\gamma^{\text{THG}}_{xxxx}$}  & & \multicolumn{4}{c}{$\gamma^{\text{THG}}_{yyyy}$}  \vspace{0.05cm}\\ 
 \cline{3-6}  \cline{8-11}  \vspace{-0.2cm} \\
& PW(3) & $-1750$  & $-2000$  & $-2320$  & $-796$   &  & $-546$  & $-562$  & $-653$  & $-14.1$   &  \\ 
& LRCW(2,1) & $-1980$  & $-1930$  & $-2390$  & $-866$   &  & $-475$  & $-1040$  & $-766$  & $-174$   &  \\ 
& QRCW(2,1) & $-1820$  & $-2050$  & $-2380$  & $-831$   &  & $-559$  & $-584$  & $-668$  & $-23.2$   &  \\ 
 & Response  & $-1800$  & $ $  & $-2380$  & $-820$   &  & $-565$  & $ $  & $-675$  & $-13.8$   &  \vspace{0.2cm} \\ 
 & & \multicolumn{4}{c}{$\gamma^{\text{THG}}_{zzzz}$} \vspace{0.05cm} \\ 
  \cline{3-6}  \vspace{-0.2cm} \\
& PW(3) & $-1040$  & $-1150$  & $-1320$  & $-469$   &  \\ 
& LRCW(2,1) & $-1130$  & $-1050$  & $-1310$  & $-531$   &  \\ 
& QRCW(2,1) & $-1090$  & $-1180$  & $-1360$  & $-491$   &  \\ 
 & Response  & $-1080$  & $ $  & $-1360$  & $-483$   & \vspace{0.2cm}  \\ 
Ne  & & \multicolumn{4}{c}{$\gamma^{\text{THG}}_{jjjj}$} &  \quad  CH$_4$  \quad & \multicolumn{4}{c}{$\gamma^{\text{THG}}_{jjjj}$} \vspace{0.05cm} \\ 
 \cline{3-6}  \cline{8-11}  \vspace{-0.2cm} \\
& PW(3) & $-119$  & $-129$  & $-148$  & $-61.3$   &  & $-2670$  & $-2750$  & $-2910$  & $19.2$   &  \\ 
& LRCW(2,1) & $-74.6$  & $-189$  & $-120$  & $-63.5$   &  & $-3950$  & $-3020$  & $-4140$  & $-720$   &  \\ 
& QRCW(2,1) & $-118$  & $-128$  & $-153$  & $-65.5$   &  & $-2650$  & $-2800$  & $-2910$  & $2.03$   &  \\ 
 & Response  & $-122$ & $ $ & $-151$ & $-62.8$  &  & $-2720$  & $ $  & $-2960$  & $46.0$  \vspace{0.2cm} \\ 
\hline 
\hline
\end{tabular}
\end{table*}

\begin{table*}
\caption{The degenerate four wave mixing component $\gamma^{\text{DFWM}}_{jjjj}$ of the second hyperpolarizabilities of Ne, HF, H$_2$O, NH$_3$ and CH$_4$ extracted from TDCCSD, TDOMP2, TDCC2, and TDCIS simulations
    compared with results from cubic response theory. \label{sec_pol_tab2}}
\centering
\begin{tabular}{l l r r r r r r r r r r  }
\hline 
\hline
\vspace{-0.1cm} \\
&  & TDCCSD & TDOMP2 & TDCC2 & TDCIS &  &  TDCCSD & TDOMP2 & TDCC2 & TDCIS \\ 
\hline
\vspace{-0.2cm}\\
HF  & & \multicolumn{4}{c}{$\gamma^{\text{DFWM}}_{xxxx}$}  & & \multicolumn{4}{c}{$\gamma^{\text{DFWM}}_{zzzz}$} \vspace{0.05cm}  \\ 
 \cline{3-6}  \cline{8-11} \vspace{-0.2cm} \\
& PW(3) & $-254$  & $-284$  & $-342$  & $-109$   &  & $-303$  & $-312$  & $-397$  & $-184$   &  \\ 
& LRCW(2,1) & $-326$  & $-360$  & $-397$  & $-134$   &  & $-354$  & $-334$  & $-424$  & $-219$   &  \\ 
& QRCW(2,1) & $-290$  & $-324$  & $-393$  & $-128$   &  & $-356$  & $-365$  & $-439$  & $-215$   &  \\ 
 & Response  & $-298$  & $ $  & $-387$  & $-128$   &  & $-358$  & $ $  & $-441$  & $-217$   &  \\ 
NH$_3$  & & \multicolumn{4}{c}{$\gamma^{\text{DFWM}}_{yyyy}$}  & & \multicolumn{4}{c}{$\gamma^{\text{DFWM}}_{zzzz}$}  \vspace{0.05cm} \\ 
 \cline{3-6}  \cline{8-11} \vspace{-0.2cm} \\
& PW(3) & $-913$  & $-955$  & $-1050$  & $64.2$   &  & $-4940$  & $-5550$  & $-6100$  & $-2690$   &  \\ 
& LRCW(2,1) & $-948$  & $-1310$  & $-1430$  & $-71.9$   &  & $-5900$  & $-6540$  & $-7350$  & $-3240$   &  \\ 
& QRCW(2,1) & $-1100$  & $-1140$  & $-1260$  & $73.9$   &  & $-5930$  & $-6570$  & $-7230$  & $-3190$   &  \\ 
 & Response  & $-1090$  & $ $  & $-1260$  & $78.8$   &  & $-5870$  & $ $  & $-7240$  & $-3200$   &  \\ 
H$_2$O  & & \multicolumn{4}{c}{$\gamma^{\text{DFWM}}_{xxxx}$}  & & \multicolumn{4}{c}{$\gamma^{\text{DFWM}}_{yyyy}$} \vspace{0.05cm}  \\ 
 \cline{3-6}  \cline{8-11} \vspace{-0.2cm} \\
& PW(3) & $-1260$  & $-1420$  & $-1640$  & $-590$   &  & $-438$  & $-449$  & $-521$  & $5.67$   &  \\ 
& LRCW(2,1) & $-1550$  & $-1660$  & $-1950$  & $-715$   &  & $-495$  & $-681$  & $-651$  & $-44.3$   &  \\ 
& QRCW(2,1) & $-1510$  & $-1690$  & $-1950$  & $-706$   &  & $-522$  & $-538$  & $-621$  & $4.39$   &  \\ 
 & Response  & $-1510$  & $ $  & $-1950$  & $-703$   &  & $-524$  & $ $  & $-623$  & $7.44$   &  \\ 
 & & \multicolumn{4}{c}{$\gamma^{\text{DFWM}}_{zzzz}$} \vspace{0.05cm}  \\ 
  \cline{3-6} \vspace{-0.2cm} \\
& PW(3) & $-803$  & $-877$  & $-1010$  & $-349$   &  \\ 
& LRCW(2,1) & $-975$  & $-1010$  & $-1180$  & $-431$   &  \\ 
& QRCW(2,1) & $-963$  & $-1050$  & $-1200$  & $-419$   &  \\ 
 & Response  & $-959$  & $ $  & $-1200$  & $-417$   &  \\ 
Ne  & & \multicolumn{4}{c}{$\gamma^{\text{DFWM}}_{jjjj}$} &  \quad CH$_4$  \quad & \multicolumn{4}{c}{$\gamma^{\text{DFWM}}_{jjjj}$}  \vspace{0.05cm} \\ 
 \cline{3-6}  \cline{8-11} \vspace{-0.2cm} \\
& PW(3) & $-79.3$  & $-84.7$  & $-95.7$  & $-41.8$   &  & $-1730$  & $-1750$  & $-1870$  & $313$   &  \\ 
& LRCW(2,1) & $-80.6$  & $-114$  & $-105$  & $-51.3$   &  & $-2330$  & $-2130$  & $-2500$  & $192$   &  \\ 
& QRCW(2,1) & $-93.1$  & $-99.7$  & $-114$  & $-50.4$   &  & $-2030$  & $-2080$  & $-2200$  & $373$   &  \\ 
 & Response  & $-94.2$ & $ $ & $-114$ & $-49.7$  &  & $-2050$  & $ $  & $-2210$  & $384$ \vspace{0.2cm} \\ 
\hline 
\hline
\end{tabular}
\end{table*}

The errors of the first hyperpolarizabilities presented in this section can be compared with the goodnesses of fit ($r^2$) of the signals they were extracted from. 
Taking NH$_3$ as an example, increasing $n_r$ from 1 to 7 yields the errors (4.68, 1.08, 1.50, 0.53, 0.93, 0.34, 0.65)\% with the LRCW method.
As expected, the errors correlate strongly to the coefficient of determination of the signal they where extracted from; the $r^2$ value
increases as (0.640, 0.989,  0.938, 0.997, 0.977, 0.999, 0.988). This correlation holds true for all molecules, confirming that the $r^2$ value can be used as a crude indicator of accuracy.


\subsection{Second hyperpolarizability}

Accurate extraction of second hyperpolarizabilies is well known to be challenging, showing errors an order of magnitude greater than those observed for the first hyperpolarizabilities.~\cite{ding}
An increase of the $r^2$ values in the third-order dipole response signal will thus substantially increase trust in second hyperpolarizabilies extracted from real-time simulations.

The errors of the $\gamma_{zzzz}^{\text{THG}}$ components as functions of simulation time are compared for the different approaches in Fig.~\ref{sec_hyp_thg}
at the CCSD level. Results for the other Cartesian components can be found in the supplementary material, including results obtained from simulations at other
levels of theory.


Starting with two cycles of simulation time, the LRCW(1,1) approach produces $\gamma_{zzzz}^{\text{THG}}$ results with errors that vary greatly between
the systems, spanning from $-1.5\%$ for H$_2$O to $+56\%$ for Ne. Increasing the ramping time to seven optical cycles (LRCW(7,1)) does not reduce the errors
which now range from $-34\%$ for CH$_4$ to $-16\%$ for Ne.
The lacking improvement highlights the convergence issues of the LRCW approach, as is clearly visible in the  CH$_4$ and H$_2$O panels of Fig. \ref{sec_hyp_thg}.
The poor convergence is accompanied by dipole response signals $\mu^{(3)}_{zzz}(t)$ that do not show the expected form. 
For example, the coefficient of determination for CH$_4$ ranges from $r^2 = 0.09$ to $r^2 = 0.22$ as the number of ramping cycles is increased.

The PW(2) approach produces $\gamma_{zzzz}^{\text{THG}}$ with small intersystem variations in error; the smallest error is 13\% for HF, the largest $17.8 \%$ for H$_2$O. The properties also converge in a systematic fashion with increasing simulation time. At the maximum simulation time, the PW(8) method is able to
attain errors as small as $-0.02 \%$ for HF and no larger than $-0.25 \%$ for H$_2$O. 

The QRCW approach produces $\gamma_{zzzz}^{\text{THG}}$ results with the smallest errors. The improvement over the other approaches is most striking when only a few optical cycles of simulation time are used. With QRCW (1,1), one gets $\gamma_{zzzz}^{\text{THG}}$ results with errors smaller than $1.6\%$ for HF, H$_2$O and NH$_3$ and somewhat higher errors for Ne and CH$_4$. Including an extra ramping cycle, the QRCW(2,1) method reduces the errors of Ne from
$-17\%$ to $3\%$ and of CH$_4$ from $-21\%$ to $2.0\%$. Hence, the QRCW(2,1) method stands out as a possible compromise between computational cost and accuracy.
The most accurate $\gamma_{zzzz}^{\text{THG}}$ results are acquired using the QRCW(7,1) method, achieving accuracies with errors below $0.8\%$ for all systems. 
However, the lowest error obtained for $\gamma_{zzzz}^{\text{THG}}$ is still a thousandfold greater than that found for the polarizabilities and $\beta_{zzzz}^{\text{OR}}$, showing that some loss of accuracy must be expected for higher-order responses.

The extracted $\gamma_{zzzz}^{\text{DFWM}}$ results behave similarly to $\gamma_{zzzz}^{\text{THG}}$. The errors are generally large when two cycles of simulation time is used,  $20\%$ for LRCW(1,1), $31\%$ for PW(2), and $-6.7\%$ for QRCW(1,1).
Increasing the simulation time quickly leads to very accurate results when using the QRCW approach, less so for the PW approach, and the LRCW approach continues to perform irregularly as a function of simulation time.
The main difference between the  $\gamma_{zzzz}^{\text{DFWM}}$ and the $\gamma_{zzzz}^{\text{THG}}$ results lies in the slower convergence with increasing simulation time for the PW approach. The inferior performance of PW for $\gamma_{zzzz}^{\text{DFWM}}$ resembles the poorer performance found for $\beta^{\text{SHG}}$.
The $\gamma_{zzzz}^{\text{DFWM}}$ results are accurate to within $0.3\%$ for all systems when using the QRCW(7,1) approach.

Also, for the second hyperpolarizability we find a relation between $r^2$ values of the fit with the errors observed, confirming that $r^2$ values close to $1$
are needed for a reliable extraction.
The results obtained with the TDCCSD, TDOMP2, TDCC2, and TDCIS methods for $\gamma^{\text{THG}}_{xxxx}$, $\gamma^{\text{THG}}_{yyyy}$, and $\gamma^{\text{THG}}_{zzzz}$ after a total simulation time of three optical cycles are shown in Table \ref{sec_pol_tab}. The $\gamma^{\text{DFWM}}_{xxxx}$, $\gamma^{\text{DFWM}}_{yyyy}$, and $\gamma^{\text{DFWM}}_{zzzz}$ results are shown in Table \ref{sec_pol_tab2}.
In general, we observe that the QRCW(2,1) approach yields more accurate second hyperpolarizabilities than the LRCW(2,1) and PW(3) approaches regardless of
the electronic-structure method used.
We note in passing that, in agreement with the observations made in Ref.~\citen{omp2},
the TDOMP2 method yields optical properties that fall between those of the TDCC2 and TDCCSD methods.
Although there is no response data available for the OMP2 method, we may assume that the TDOMP2 values reported with the QRCW approach
is correct to within $1\%$ based on the accuracies observed for the TDCCSD, TDCC2, and TDCIS methods.

As expected,~\cite{Larsen1999} the simulations at the TDCIS level provides optical properties vastly different from the other three methods due to lack of electron correlation.
Since the fourth-order
dipole signal is very weak, separation by numerical differentiation is more challenging.
Although the optimal choice of the electric-field strength is beyond the scope of this paper,
we remark that increasing the electric-field strength from $E= 0.001$ to $E= 0.004$ seems to reduce the error of the numerical differentiation for the TDCIS
method.

\begin{figure}
\includegraphics[scale=0.55]{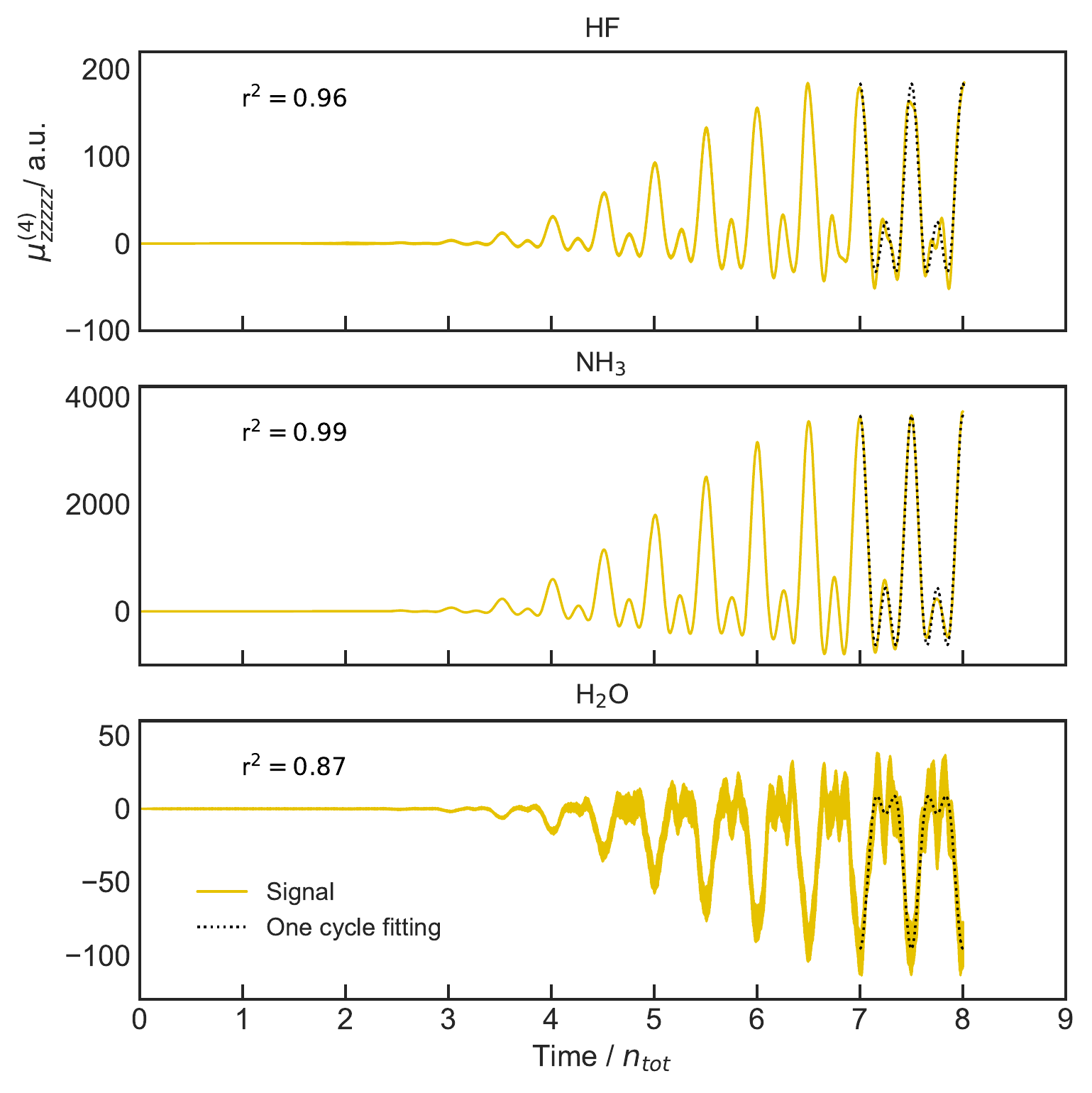} 
    \caption{The fourth-order dipole response functions obtained from TDCCSD simulations for the HF, NH$_3$ and H$_2$O molecules using QRCW(7,1).
    \label{fifth_order_fitting} }
\end{figure}


\begin{table}
\centering
    \caption{The third hyperpolarizabilities $\delta_{zzzzz}$ extracted from TDCCSD simulations with QRCW(2,1), QRCW(7,1) , PW(3) and PW(8).
    \label{fourth_order_values}}
\begin{tabular}{l l r r r r r r  r r r r r r }
\hline 
\hline
\vspace{-0.2cm} \\
 & &  $\delta_{zzzzz}^{\text{FHG}}$ &  $\delta_{zzzzz}^{\text{HSHG}}$  &  $\delta_{zzzzz}^{\text{HOR}}$   &  \\ 
 HF \\
 &  QRCW(2,1)  &  $10\,636$  &  $3\,863$ &  $2\,726$  \\
&  QRCW(7,1)  &  $11\,948$  &  $3\,808$ &  $2\,705$  \\
&  PW(3)  &   $14\,795$  &  $3\,082$ & $2\,641$ \\
&  PW(8)  &   $12\,512$  &  $3\,699$ & $2\,724$ \\
%
NH$_3$ &  \\	 
& QRCW(2,1)& $324\,317$ &  $76\,998$ & $56\,012$\\
& QRCW(7,1)&\quad $231\,482$ & \quad $81\,326$ & \quad $53\,906$\\
&  PW(3)    &$244\,356$  & $62\,355$  &  $52\,571$\\
&  PW(8)    &$240\,228$  & $74\,587$  &  $54\,103$\\
%
 H$_2$O & &    \\ 
& QRCW(2,1)  & $-5\,067$  &  $-2\,271$   & $-1\,694$   \\
& QRCW(7,1)   & $-4\,520$  &  $-2\,208$   & $-1\,662$   \\
&  PW(3)    &$-4\,452$ & $-1\,748$ & $-1\,608$  \\
&  PW(8)    &$-4\,536$ & $-2\,129$ & $-1\,665$  \\
\hline
\hline
\end{tabular}
\end{table}

\subsection{Third hyperpolarizability}

The RCW and PW approaches can straightforwardly be extended to evaluate higher order properties, here exemplified with third hyperpolarizabilites. An example of extracting fourth hyperpolarizabilities can be found in the supplementary material.

To the best of our knowledge, there are no implemented analytic response functions available for third hyperpolarizabilities.
Instead, we will use the $r^2$ value of the fit to the fourth-order dipole response function $\mu^{(4)}(t)$ to gauge the accuracy of the extracted
third hyperpolarizabilities. 
Figure \ref{fifth_order_fitting} shows $\mu^{(4)}_{zzzzz}(t)$ calculated at the TDCCSD level of theory along with
the associated curve fitting using the QRCW(7,1) method.
The fitted curves have $r^2$ values ranging from $0.87$ to $0.99$, indicating that only the third hyperpolarizability results for NH$_3$ ($r^2 = 0.99$)
and HF ($r^2 = 0.96$) are fully reliable while the results for H$_2$O are decent estimates.

The third hyperpolarizabilities for HF,  NH$_3$, and H$_2$O are given in in Table \ref{fourth_order_values}. The discrepancies between the PW and RCW approaches are modest, typically on the order of $3 \%$ for the $\delta_{zzzzz}^{\text{FHG}}$ and $0.5 \%$  for the  $\delta_{zzzzz}^{\text{HSHG}}$ when simulating for a total of eight optical cycles.
Increasing the total simulation time (i.e., increasing $n_r$ in the case of the QRCW approach) is likely to improve the accuracy further, in analogy to
the convergence behavior observed for the lower-order responses above.

\section{Concluding remarks}
\label{sec:concluding}

We have compared three approaches to the extraction of linear and nonlinear optical properties from electron dynamics simulations with respect to 
accuracy relative to results from response theory and computational effort. The LRCW and QRCW approaches are based on a monochromatic continuous wave
perturbation ramped from zero to full strength linearly and quadratically, while the PW approach uses Fourier filtering to extract properties at a given
frequency from signals recorded during the interaction of electrons with a finite laser pulse. All three approaches rely on numerical (finite-difference)
differentiation to separate different orders of response in the time domain, followed by curve fitting to obtain the property of interest at a given
frequency.

Showing irregular convergence behavior towards response results as the ramping time is increased, the LRCW approach is difficult to apply reliably
for higher-order nonlinear responses. Using a single optical cycle for ramping, as previously recommended,~\cite{ding} is found to be insufficient.
On the other hand, we find that the post-ramp simulation time can be reduced to a single cycle without incurring an accuracy penalty, as opposed to 
the four ramping cycles proposed by \citeauthor{ding}.~\cite{ding}

The QRCW approach, proposed in this work, is a clear improvement due to reduced nonadiabatic effects.
Although the convergence behavior remains somewhat irregular, errors observed for linear and nonlinear response properties are significantly reduced.
Our tests indicate that the QRCW approach yields highly accurate linear and quadratic response properties with simulation times as short as two optical
cycles, one cycle for ramping and one post-ramp cycle for extracting the response property of interest. An additional ramp cycle should be added to reliably
extract cubic response properties, however. The QRCW approach thus yields significantly improved accuracy at about half the computational cost of
the LRCW approach as proposed by \citeauthor{ding}.~\cite{ding}

We find that the coefficient of determination ($r^2$) obtained for the curve fitting can be used as an indicator of accuracy in lieu of analytical results from
response theory. In all cases studied in this work, the $r^2$ value can be improved by increasing the ramping time of the QRCW approach.

The PW approach offers an alternative to QRCW. The PW approach shows monotonous but typically rather slow convergence towards response results
with respect to simulation time. For comparable accuracy, the PW approach typically requires much longer simulation times than the QRCW approach,
which we recommend for reliable and efficient extraction of linear and nonlinear response properties.

While our tests are based on TDCC and TDCIS methods, we expect that our conclusions remain valid also for other electronic-structure methods
such as real-time time-dependent density-functional theory.

\section*{Supplementary material}
 
The HF, H$_2$O, NH$_3$, and CH$_4$ geometries used throughout this article can be found in section \rom{1} of the supplementary material.
The procedure for finding the fifth-order hyperpolarizabilities and example calculations for the HF molecule are found in section \rom{2}.
Tables displaying polarizabilites, first hyperpolarizabilites, and second hyperpolarizabilites extracted using QRCW(7,1), LRCW(7,1), and PW(8)
for all unique diagonal directions at the CC2, OMP2, CIS, and CCSD levels of theory are available in section \rom{3}.
Finally, figures displaying the relative errors of the polarizabilites, first hyperpolarizabilites, and second hyperpolarizabilites as functions
of $n_{tot} = (2,3,4,5,6,7,8)$ at the CC2 and CCSD levels of theory for all diagonal directions are available in section \rom{4}.

\section*{Acknowledgment}

This work was supported by the Research Council of Norway through its Centres of Excellence scheme, project number 262695.
The calculations were performed on resources provided by Sigma2---the National Infrastructure for High Performance Computing and
Data Storage in Norway, Grant No.~NN4654K.
S.~K.~and T.~B.~P.~acknowledge the support of the Centre for Advanced Study in Oslo, Norway, which funded and hosted the
CAS research project \emph{Attosecond Quantum Dynamics Beyond the Born-Oppenheimer Approximation} during the academic year
2021-2022.

\section*{Data availability statement}

The data that support the findings of this study are available from the corresponding author upon reasonable request.

\bibliography{Compare_pulse_ramp}

\end{document}


\newgeometry{top=30mm, bottom=25mm} 
\title{Supplementary material for ``Adiabatic extraction of nonlinear optical properties from real-time time-dependent electronic-structure theory''}
\author{Benedicte Sverdrup Ofstad}
\email{b.s.ofstad@kjemi.uio.no}
\affiliation{Hylleraas Centre for Quantum Molecular Sciences, Department of Chemistry,
             University of Oslo, Norway}
\author{H{\aa}kon Emil Kristiansen}
\affiliation{Hylleraas Centre for Quantum Molecular Sciences, Department of Chemistry,
             University of Oslo, Norway}
\author{Einar Aurbakken}
\affiliation{Hylleraas Centre for Quantum Molecular Sciences, Department of Chemistry,
             University of Oslo, Norway}
\author{{\O}yvind Sigmundson Sch{\o}yen}
\affiliation{Department of Physics, University of Oslo, Norway}
\author{Simen Kvaal}
\affiliation{Hylleraas Centre for Quantum Molecular Sciences, Department of Chemistry,
             University of Oslo, Norway}
\author{Thomas Bondo Pedersen}
\email{t.b.pedersen@kjemi.uio.no}
\affiliation{Hylleraas Centre for Quantum Molecular Sciences, Department of Chemistry,
             University of Oslo, Norway}

\date{\today}

\maketitle

\section{Molecular geometries}

\noindent The molecular geometry input used for the calculations are displayed in Table \ref{table_1}.

\begin{table}[H]
\begin{center}
\caption{Cartesian coordinates in Bohr. \label{table_1} }
\begin{tabular}{c c r r r}
\hline
\hline
& &  $x$ & $y$ & $z$ \\
Ne \\
& Ne & 0 & 0 & 0 \\
\\
HF \\
& H & 0 & 0 & 0 \\
& F & 0 & 0 & $1.7328795$ \\
\\
H$_2$O \\
& O & $0$ & $0$   & $-0.1239093563$ \\
& H & $0$ & $ 1.4299372840$  & $0.9832657567$ \\
& H & $0$ & $-1.4299372840$  & $0.9832657567$ \\
\\
NH$_3$ \\
&N & $0$    & $ 0$    & $ 0.2010$ \\
&H & $0$    & $ 1.7641$ & $-0.4690$ \\
&H & $1.5277$ & $-0.8820$ & $-0.4690$ \\
&H & $-1.5277$& $-0.8820$ & $-0.4690$ \\
\\
CH$_4$ \\
& C & $0$ & $0$ & $0$ \\ 
& H & $1.2005$ & $1.2005$ & $1.2005$ \\ 
& H & $-1.2005$ & $-1.2005$ & $1.2005$ \\
& H & $-1.2005$ & $1.2005$ & $-1.2005$ \\
& H & $1.2005$ & $-1.2005$ & $-1.2005$  \\
\hline
\hline
\end{tabular}
\end{center}
\end{table}

\section{Fourth hyperpolarizability}

\noindent To obtain the fourth hyperpolarizability, the time-dependent dipole moment is expanded to fifth order
\begin{align}
    \mu_i(t) &= \mu_i(0) + \sum_j \mu_{ij}^{(1)}(t) \mathcal{E}_j
    + \sum_{jk} \mu_{ijk}^{(2)}(t) \mathcal{E}_j \mathcal{E}_k
    + \sum_{jkl} \mu_{ijkl}^{(3)}(t) \mathcal{E}_j \mathcal{E}_k  \mathcal{E}_l
    \nonumber \\
    &+ \sum_{jklm} \mu_{ijklm}^{(4)}(t) \mathcal{E}_j \mathcal{E}_k  \mathcal{E}_l \mathcal{E}_m +  \sum_{jklmn} \mu_{ijklmn}^{(5)}(t) \mathcal{E}_j \mathcal{E}_k  \mathcal{E}_l \mathcal{E}_m \mathcal{E}_n
    + \cdots.
\end{align}
The central difference formula for extracting $\mu_{ijjjjj}^{(5)}(t)$ is
\begin{align}
\label{central_diff}
    \mu_{ijjjjj}^{(5)}(t)
    &=
    \frac{29\Delta_i^-(t;\mathcal{E}_j) - 26\Delta_i^-(t;2\mathcal{E}_j) +9 \Delta_i^-(t;3\mathcal{E}_j)-\Delta_i^-(t;4\mathcal{E}_j)}{720\mathcal{E}_j^5}
    + O(\mathcal{E}_j^4).
\end{align}
The quintic response is given by
\begin{align}
  \mu_{ijjjjj}^{(5)}(t) = \frac{1}{120}& \iiint\!\!\!\iint_{-\infty}^\infty  \epsilon_{ijjjjj}(-\omega^{(5)}; \omega_1, \omega_2, \omega_3, \omega_4, \omega_5) \nonumber \\
          &\times \tilde{F}(\omega_1)\tilde{F}(\omega_2)\tilde{F}(\omega_3)\tilde{F}(\omega_4)\tilde{F}(\omega_5) \ee^{-\ii (\omega_1 + \omega_2 + \omega_3 + \omega_4 + \omega_5) t} \nonumber \\ 
        &\times \text{d}\omega_1\text{d}\omega_2\text{d}\omega_3\text{d}\omega_4 \text{d}\omega_5 .
\end{align}
With $F(t)= \text{cos}(\omega t)$, one obtains
\begin{align}
    \mu_{ijjjj}^{(5)}(t) &= \frac{1}{1920}
    \big[
         \epsilon_{ijjjjj}(-5\omega; \omega, \omega, \omega, \omega, \omega) \cos(5\omega t)
    \nonumber \\
    &\;\;\;\;\;\;
     +5\epsilon_{ijjjjj}(-3\omega; \omega, \omega, \omega, \omega, -\omega) \cos(3\omega t)
    \nonumber \\
    &\;\;\;\;\;\;
     +10\epsilon_{ijjjjj}(-\omega; \omega, \omega, \omega, -\omega, -\omega) \cos(\omega t)  
    \big]
    \nonumber \\
    &=
    \big[
        \epsilon^\text{5HG}_{ijjjjj}(\omega) \cos(5\omega t)
    \nonumber \\
    &\;\;\;\;\;\;
     +5\epsilon^\text{HTHG}_{ijjjjj}(\omega) \cos(3\omega t)
    \nonumber \\
    &\;\;\;\;\;\;
     +10\epsilon^\text{D6WM}_{ijjjjj}(\omega) \cos(\omega t)
    \big].
\label{mu5}
\end{align}

\begin{figure}[H]
\begin{center}
\begin{subfigure}{0.49\textwidth}
\includegraphics[width=\textwidth]{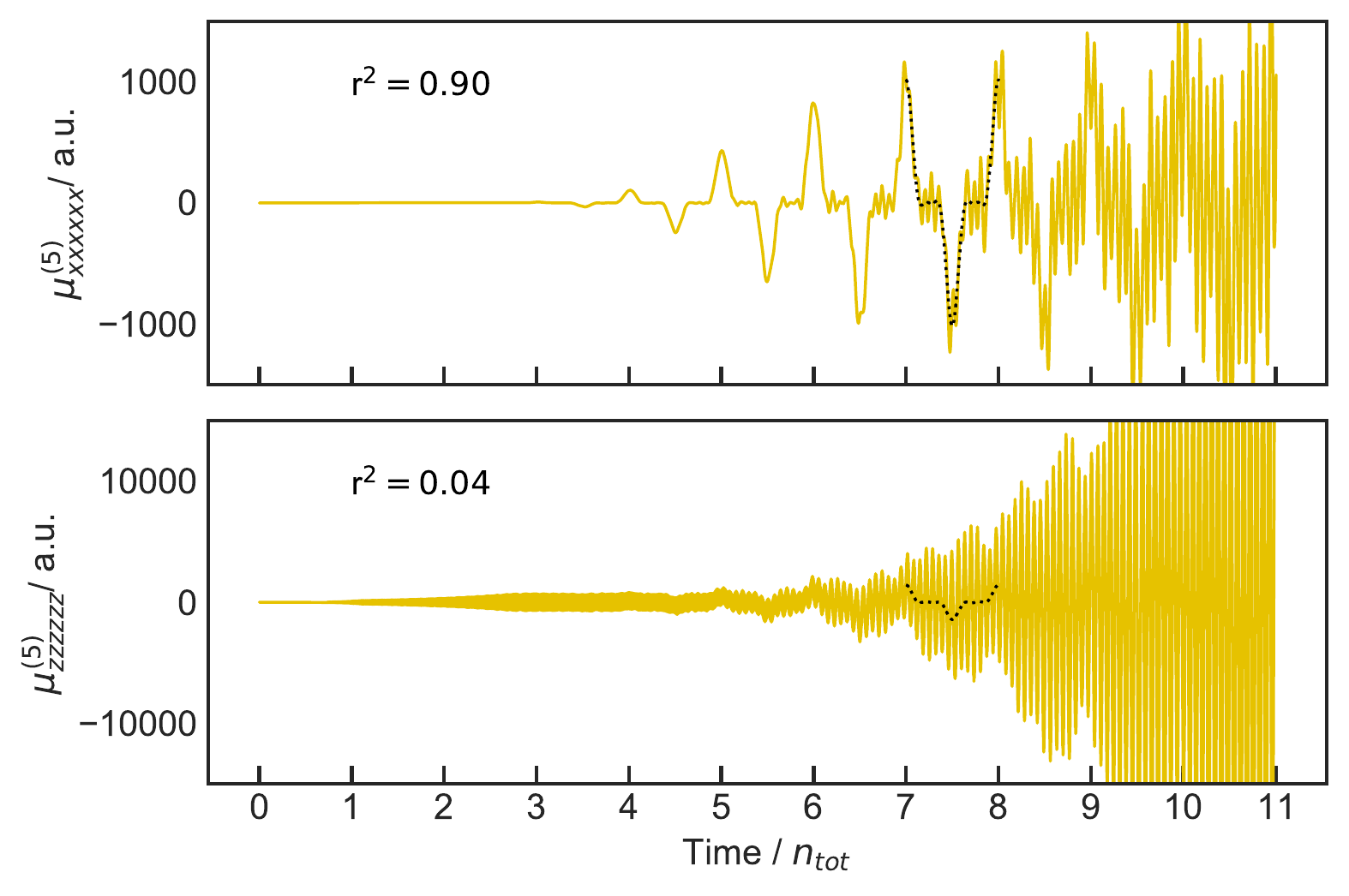}
\caption{\centering RCW}
\label{fig_1a}
\end{subfigure}
\hfill
\begin{subfigure}{0.49\textwidth}
\includegraphics[width=\textwidth]{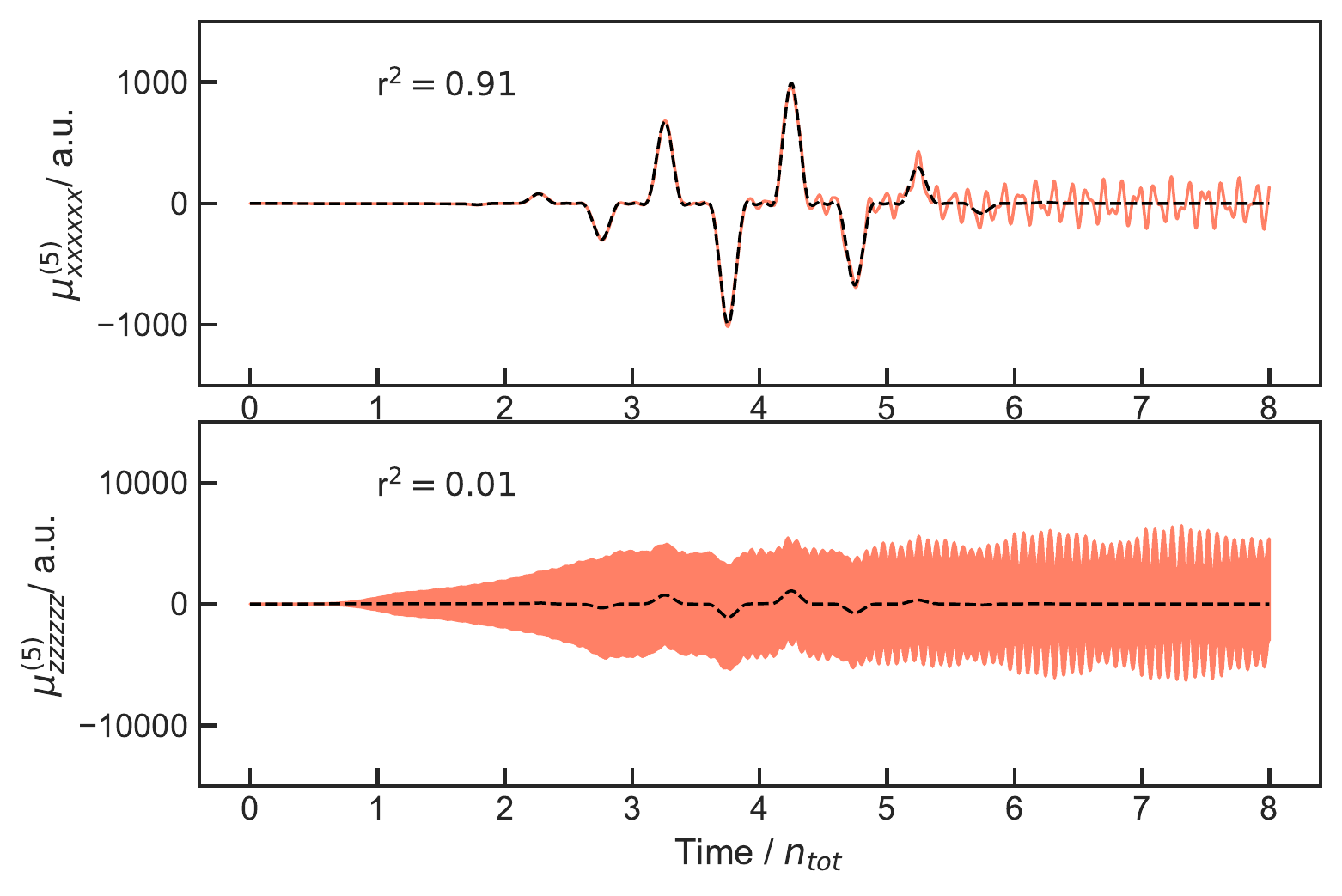}
\caption{\centering PW}
\label{fig_1b}
\end{subfigure}
\caption{The fifth-order dipole response obtained from TDCCSD simulations for HF using $n_r = 7$ quadratic ramp cycles followed by $n_p=1$ post-ramp cycles in \ref{fig_1a} and $n_{tot} = 8$ pulsed wave cycles in \ref{fig_1b}. }
\end{center}
\end{figure}

\noindent For the HF molecule, we use $\omega = 0.05$, which is below one fourth of the first excitation energy.
The fifth-order dipole signals extracted with the RCW and PW approaches are plotted in Figs.~\ref{fig_1a} and \ref{fig_1b}, respectively,
along with the least-squares fits and their $r^2$ values.
The correlation for the fourth hyperpolarizability is resonably good in the diagonal $x$-direction but very poor in the diagonal $z$-direction.
This \emph{might} be caused by instabilities in the numerical differentiation, Eq. (\ref{central_diff}), for which we have not studied convergence with
respect to field strength.
The results for the HF molecule in the $x$-direction are given in Table \ref{fourth_order_values}. The PW and QRCW values 
agree to within $8\%$.

\begin{table}
\centering
\caption{The $\epsilon^{\text{5HG}}$,  $\epsilon^{\text{HTHG}}$, and $\epsilon^{\text{D6WM}}$ component in the diagonal $x-$ and $y-$ direction of the fourth hyperpolarizability is extracted for the HF molecule at the CCSD level \label{fourth_order_values}}
\begin{tabular}{l l r r r r r r  r r r r r r }
\hline 
\hline
\vspace{-0.2cm} \\
 & & \quad  $\omega$ \quad \quad & $\epsilon_{xxxxxx}^{\text{5HG}}$ & $\epsilon_{xxxxxx}^{\text{HTHG}}$ & $\epsilon_{xxxxxx}^{\text{D6WM}} $ & \quad \quad & $\epsilon_{zzzzzz}^{\text{5HG}}$ & $\epsilon_{zzzzzz}^{\text{HTHG}}$ & $\epsilon_{zzzzzz}^{\text{D6WM}} $\\ 
HF\\
&  QRCW(7,1) &  \quad 0.05 \quad \quad &216238& 135160 & 106098 & \quad  & 408 000 & 183 000 & 142 000  \\
&  PW(8)          &  \quad 0.05 \quad \quad& 234506 &137296 & 102577 &\quad   & 203 365 & 146 188& 119 526 \\
\vspace{-0.2cm} \\
\hline
\hline
\end{tabular}
\end{table}

\section{Polarizabilities and first hyperpolarizabilities from TDCCSD, TDOMP2, TDCC2, and TDCIS simulations}

All unique diagonal cartesian components are presented for the polarizabilites in Table \ref{pol} and for the first hyperpolarizabilites in Table \ref{hyppol} after a total of two optical cycles for the LRCW(1,1), QRCW(1,1) and PW(2) approach.

\begin{table}[H]
\centering
\caption{Polarizabilites  \label{pol}}
\begin{tabular}{l l r r r r r r r r r r  }
\hline 
\hline
\vspace{-0.1cm} \\
&  & TDCCSD & TDOMP2 & TDCC2 & TDCIS & &  TDCCSD & TDOMP2 & TDCC2 & TDCIS \\ 
\hline
\vspace{-0.2cm}\\
HF  & & \multicolumn{4}{c}{$\alpha_{xx}$}  & & \multicolumn{4}{c}{$\alpha_{zz}$}  \vspace{0.05cm}  \\ 
 \cline{3-6}  \cline{8-11} \vspace{-0.2cm} \\
& PW(2) & $4.4534$  & $4.5696$  & $4.7192$  & $4.2274$   &  & $6.4199$  & $6.5036$  & $6.7898$  & $6.4934$   &  \\ 
& LRCW(1,1) & $4.4450$  & $4.5657$  & $4.7210$  & $4.2238$   &  & $6.4159$  & $6.4913$  & $6.7752$  & $6.4884$   &  \\ 
& QRCW(1,1) & $4.4405$  & $4.5571$  & $4.7077$  & $4.2199$   &  & $6.4080$  & $6.4908$  & $6.7757$  & $6.4839$   &  \\ 
 & Response  & $4.4419$  & $ $  & $4.7048$  & $4.2208$   &  & $6.4076$  & $ $  & $6.7759$  & $6.4833$   &  \\ 
NH3  & & \multicolumn{4}{c}{$\alpha_{yy}$}  & & \multicolumn{4}{c}{$\alpha_{zz}$} \vspace{0.05cm}  \\ 
 \cline{3-6}  \cline{8-11} \vspace{-0.2cm} \\
& PW(2) & $13.108$  & $13.236$  & $13.567$  & $14.203$   &  & $15.062$  & $15.625$  & $15.882$  & $14.557$   &  \\ 
& LRCW(1,1) & $13.105$  & $13.233$  & $13.564$  & $14.199$   &  & $15.077$  & $15.615$  & $15.879$  & $14.553$   &  \\ 
& QRCW(1,1) & $13.101$  & $13.229$  & $13.559$  & $14.197$   &  & $15.046$  & $15.602$  & $15.860$  & $14.542$   &  \\ 
 & Response  & $13.101$  & $ $  & $13.560$  & $14.197$   &  & $15.041$  & $ $  & $15.858$  & $14.544$   &  \\ 
H2O  & & \multicolumn{4}{c}{$\alpha_{xx}$}  & & \multicolumn{4}{c}{$\alpha_{yy}$}  \vspace{0.05cm}  \\ 
 \cline{3-6}  \cline{8-11} \vspace{-0.2cm} \\
& PW(2) & $8.7901$  & $9.1625$  & $9.4244$  & $8.1642$   &  & $9.9359$  & $10.063$  & $10.436$  & $10.461$   &  \\ 
& LRCW(1,1) & $8.7941$  & $9.1534$  & $9.4151$  & $8.1665$   &  & $9.9337$  & $10.059$  & $10.433$  & $10.461$   &  \\ 
& QRCW(1,1) & $8.7810$  & $9.1544$  & $9.4161$  & $8.1602$   &  & $9.9318$  & $10.058$  & $10.431$  & $10.457$   &  \\ 
 & Response  & $8.7825$  & $ $  & $9.4150$  & $8.1599$   &  & $9.9318$  & $ $  & $10.431$  & $10.458$   &  \\ 
 & & \multicolumn{4}{c}{$\alpha_{zz}$} \vspace{0.05cm}  \\ 
  \cline{3-6}  \vspace{-0.2cm} \\
& PW(2) & $9.1164$  & $9.3398$  & $9.6315$  & $9.2544$   &  \\ 
& LRCW(1,1) & $9.1194$  & $9.3342$  & $9.6251$  & $9.2557$   &  \\ 
& QRCW(1,1) & $9.1108$  & $9.3343$  & $9.6260$  & $9.2507$   &  \\ 
 & Response  & $9.1114$  & $ $  & $9.6257$  & $9.2506$   &  \\ 
Ne  & & \multicolumn{4}{c}{$\alpha_{jj}$} &  CH4   \quad  &   \multicolumn{4}{c}{$\alpha_{jj}$} \vspace{0.05cm}  \\ 
 \cline{3-6}  \cline{8-11} \vspace{-0.2cm} \\
& PW(2) & $2.7392$  & $2.7756$  & $2.8592$  & $2.5773$   &  & $17.074$  & $17.200$  & $17.514$  & $19.104$   &  \\ 
& LRCW(1,1) & $2.7383$  & $2.7743$  & $2.8563$  & $2.5747$   &  & $17.055$  & $17.184$  & $17.495$  & $19.095$   &  \\ 
& QRCW(1,1) & $2.7363$  & $2.7724$  & $2.8558$  & $2.5751$   &  & $17.049$  & $17.176$  & $17.489$  & $19.083$   &  \\ 
 & Response  & $2.7364$ & $ $ & $2.8560$ & $2.5752$  &  & $17.050$  & $ $  & $17.490$  & $19.081$ \vspace{0.02cm}   \\ 
\hline 
\hline
\end{tabular}
\end{table}


\newgeometry{top=30mm, bottom=20mm} 

\begin{table}[H]
\caption{First hyperpolarizabilites \label{hyppol}}
\centering
\begin{tabular}{l l r r r r r r r r r r r r r }
\hline 
\hline
\vspace{-0.1cm} \\
&  & TDCCSD & TDOMP2 & TDCC2 & TDCIS & \quad \quad \quad &  TDCCSD & TDOMP2 & TDCC2 & TDCIS \\ 
\hline
\vspace{-0.2cm}\\
HF & & \multicolumn{4}{c}{$\beta_{zzz}^{\text{SHG}}$} & & \multicolumn{4}{c}{$\beta_{zzz}^{\text{OR}}$} \vspace{0.05cm}  \\ 
 \cline{3-6}  \cline{8-11}  \vspace{-0.2cm} \\
& PW(2) & $10.960$  & $11.139$  & $13.378$  & $15.121$   &  & $12.138$  & $12.311$  & $14.712$  & $17.222$   &  \\ 
& LRCW(1,1) & $14.295$  & $14.702$  & $17.832$  & $19.774$   &  & $12.816$  & $13.066$  & $15.676$  & $18.177$   &  \\ 
& QRCW(1,1) & $14.354$  & $14.613$  & $17.543$  & $19.893$   &  & $12.803$  & $12.998$  & $15.529$  & $18.210$   &  \\ 
 & Response  & $14.370$  & $ $  & $17.520$  & $19.916$   &  & $12.812$  & $ $  & $15.519$  & $18.222$   &  \\ 
NH3 & & \multicolumn{4}{c}{$\beta_{yyy}^{\text{SHG}}$} & & \multicolumn{4}{c}{$\beta_{yyy}^{\text{OR}}$} \vspace{0.05cm}  \\ 
 \cline{3-6}  \cline{8-11}  \vspace{-0.2cm} \\
& PW(2) & $-11.680$  & $-12.202$  & $-13.109$  & $-11.828$   &  & $-14.017$  & $-14.611$  & $-15.707$  & $-14.291$   &  \\ 
& LRCW(1,1) & $-15.782$  & $-16.556$  & $-17.817$  & $-15.870$   &  & $-15.032$  & $-15.700$  & $-16.894$  & $-15.274$   &  \\ 
& QRCW(1,1) & $-15.490$  & $-16.160$  & $-17.364$  & $-15.719$   &  & $-14.892$  & $-15.511$  & $-16.677$  & $-15.200$   &  \\ 
 & Response  & $-15.504$  & $ $  & $-17.397$  & $-15.714$   &  & $-14.898$  & $ $  & $-16.693$  & $-15.198$   &  \\ 
& & \multicolumn{4}{c}{$\beta_{zzz}^{\text{SHG}}$} & & \multicolumn{4}{c}{$\beta_{zzz}^{\text{OR}}$} \vspace{0.05cm}  \\ 
 \cline{3-6}  \cline{8-11}  \vspace{-0.2cm} \\
& PW(2) & $21.531$  & $27.849$  & $30.677$  & $23.919$   &  & $22.742$  & $29.020$  & $32.178$  & $27.590$   &  \\ 
& LRCW(1,1) & $29.330$  & $37.006$  & $41.003$  & $31.871$   &  & $24.518$  & $30.942$  & $34.395$  & $29.421$   &  \\ 
& QRCW(1,1) & $27.944$  & $36.114$  & $39.798$  & $31.643$   &  & $23.859$  & $30.445$  & $33.764$  & $29.271$   &  \\ 
 & Response  & $28.020$  & $ $  & $39.869$  & $31.558$   &  & $23.904$  & $ $  & $33.801$  & $29.228$   &  \\ 
H2O & & \multicolumn{4}{c}{$\beta_{zzz}^{\text{SHG}}$} & & \multicolumn{4}{c}{$\beta_{zzz}^{\text{OR}}$} \vspace{0.05cm}  \\ 
 \cline{3-6}  \cline{8-11}  \vspace{-0.2cm} \\
& PW(2) & $-7.2372$  & $-7.8969$  & $-9.9052$  & $-13.725$   &  & $-8.5816$  & $-9.3123$  & $-11.679$  & $-16.687$   &  \\ 
& LRCW(1,1) & $-9.6829$  & $-10.483$  & $-13.158$  & $-18.271$   &  & $-9.1759$  & $-9.9075$  & $-12.422$  & $-17.782$   &  \\ 
& QRCW(1,1) & $-9.5993$  & $-10.479$  & $-13.134$  & $-18.239$   &  & $-9.1164$  & $-9.8935$  & $-12.403$  & $-17.752$   &  \\ 
 & Response  & $-9.5909$  & $ $  & $-13.117$  & $-18.247$   &  & $-9.1119$  & $ $  & $-12.394$  & $-17.756$   & \vspace{0.02cm} \\ 
\hline 
\hline 
\end{tabular}
\end{table}

\section{Convergence with respect to simulation time}

\begin{figure}[H]
\begin{center}
 \subfloat[\centering TDCCSD ]{{\includegraphics[scale=0.55]{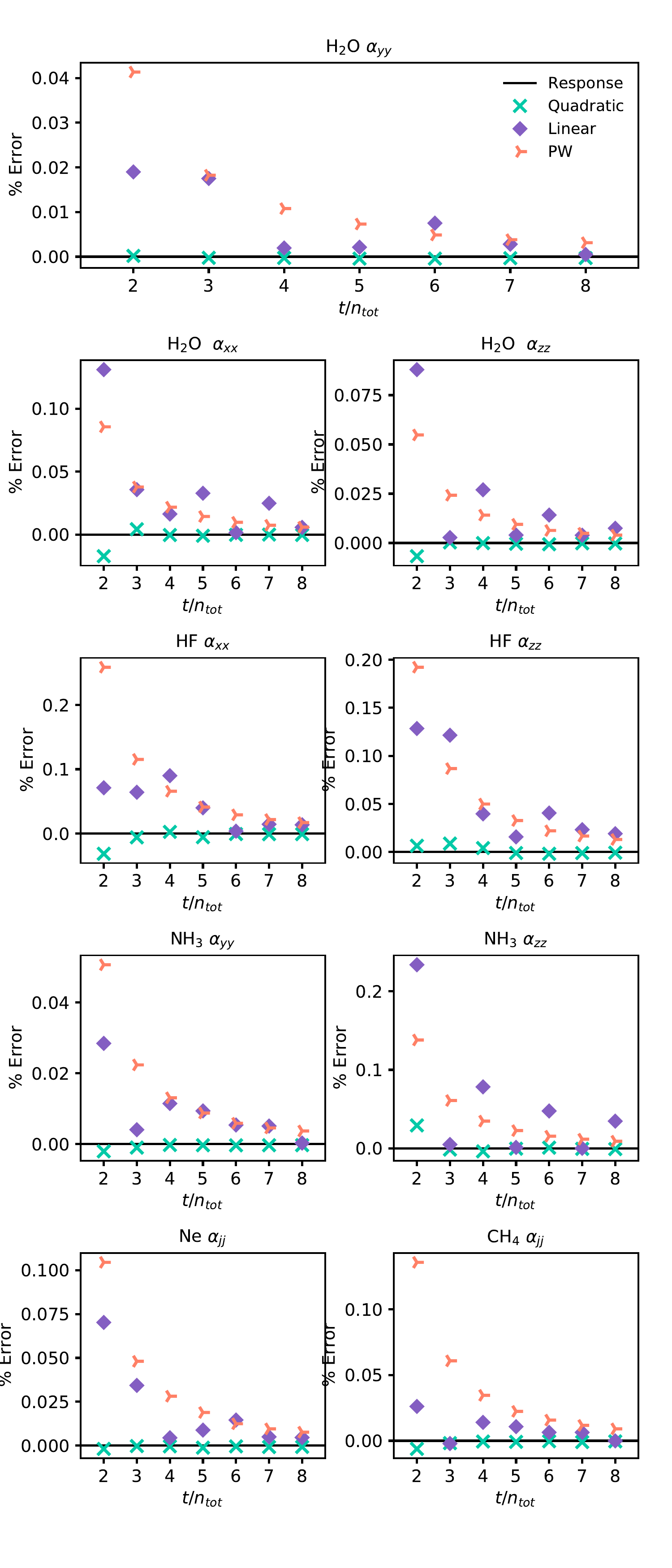} }}
 \subfloat[\centering TDCC2 ]{{\includegraphics[scale=0.55]{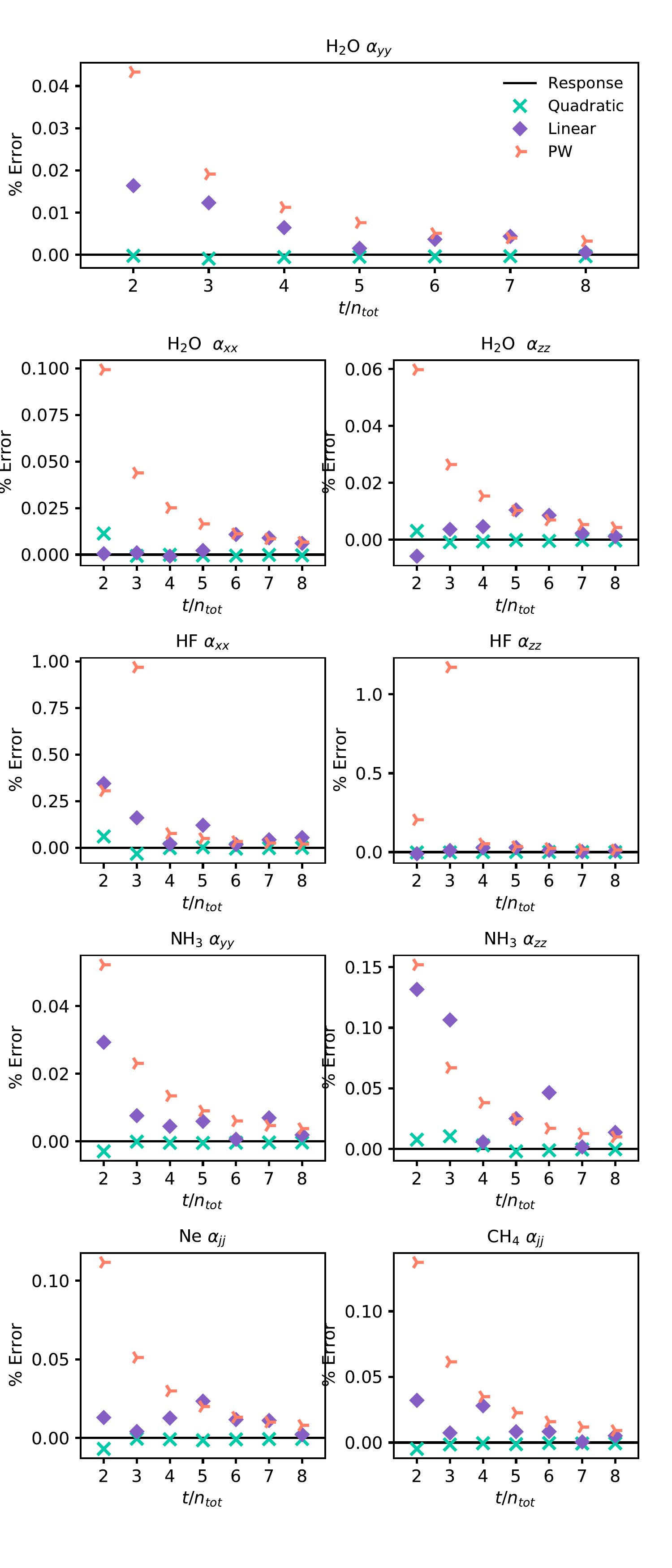} }} \\
\caption{The polarizability \label{fig1}}
\end{center}
\end{figure}

\begin{figure}[H]
\begin{center}
 \subfloat[\centering TDCCSD ]{{\includegraphics[scale=0.55]{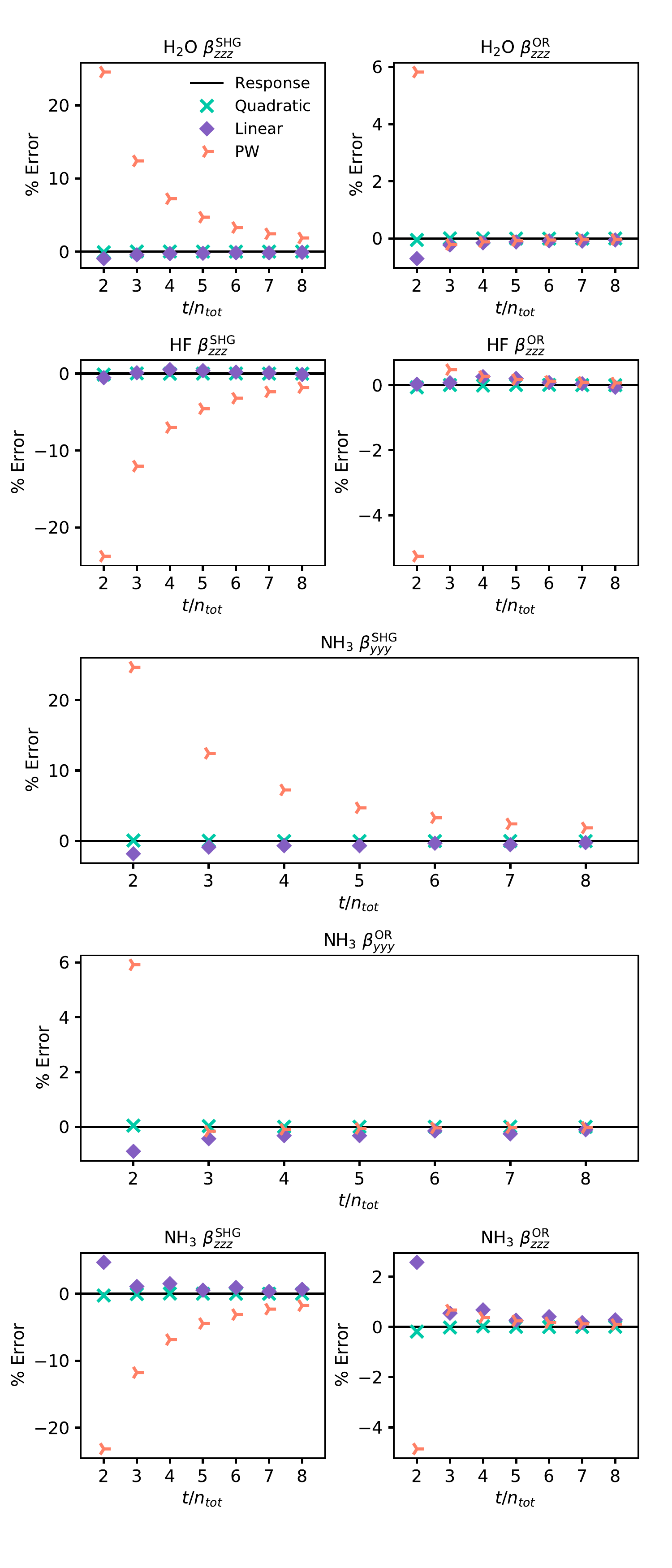} }}
 \subfloat[\centering TDCC2 ]{{\includegraphics[scale=0.55]{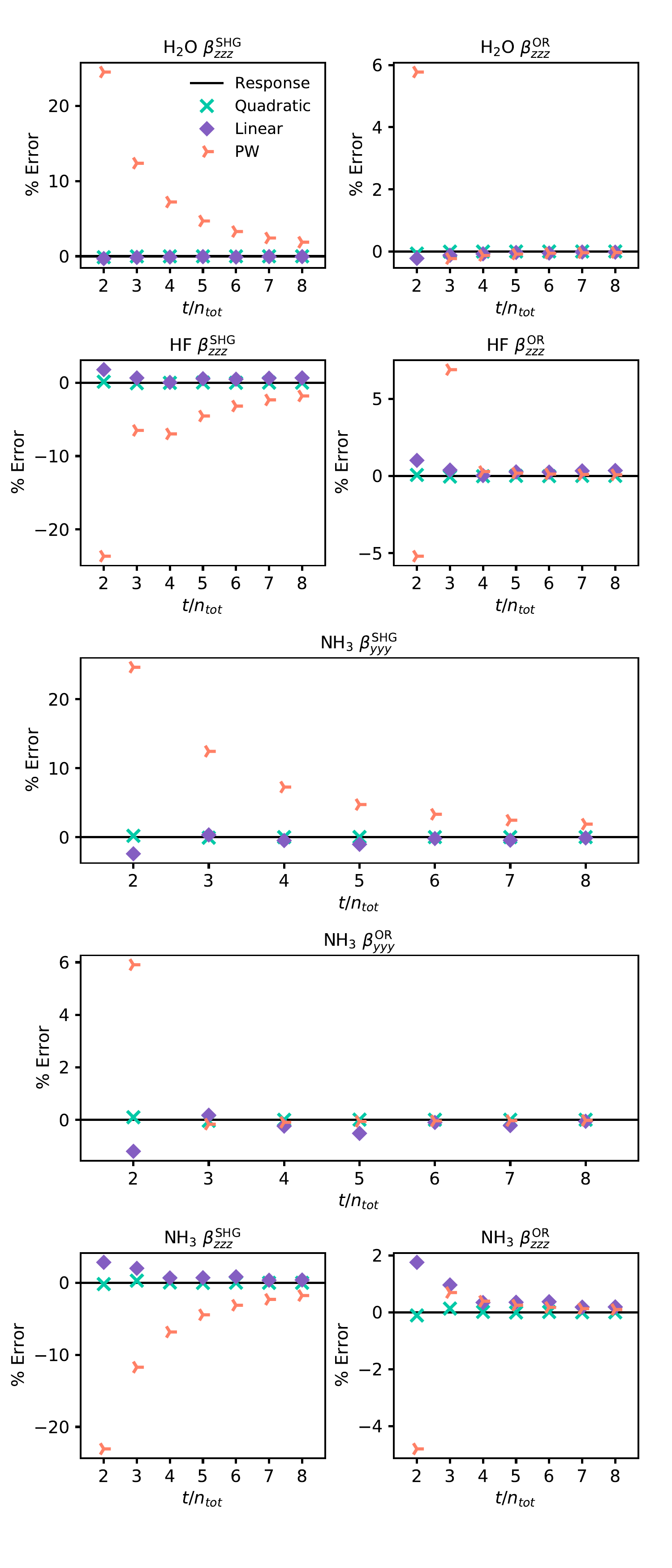} }} \\
\caption{The first hyperpolarizability \label{fig2}}
\end{center}
\end{figure}

\begin{figure}[H]
\begin{center}
 \subfloat[\centering TDCCSD ]{{\includegraphics[scale=0.55]{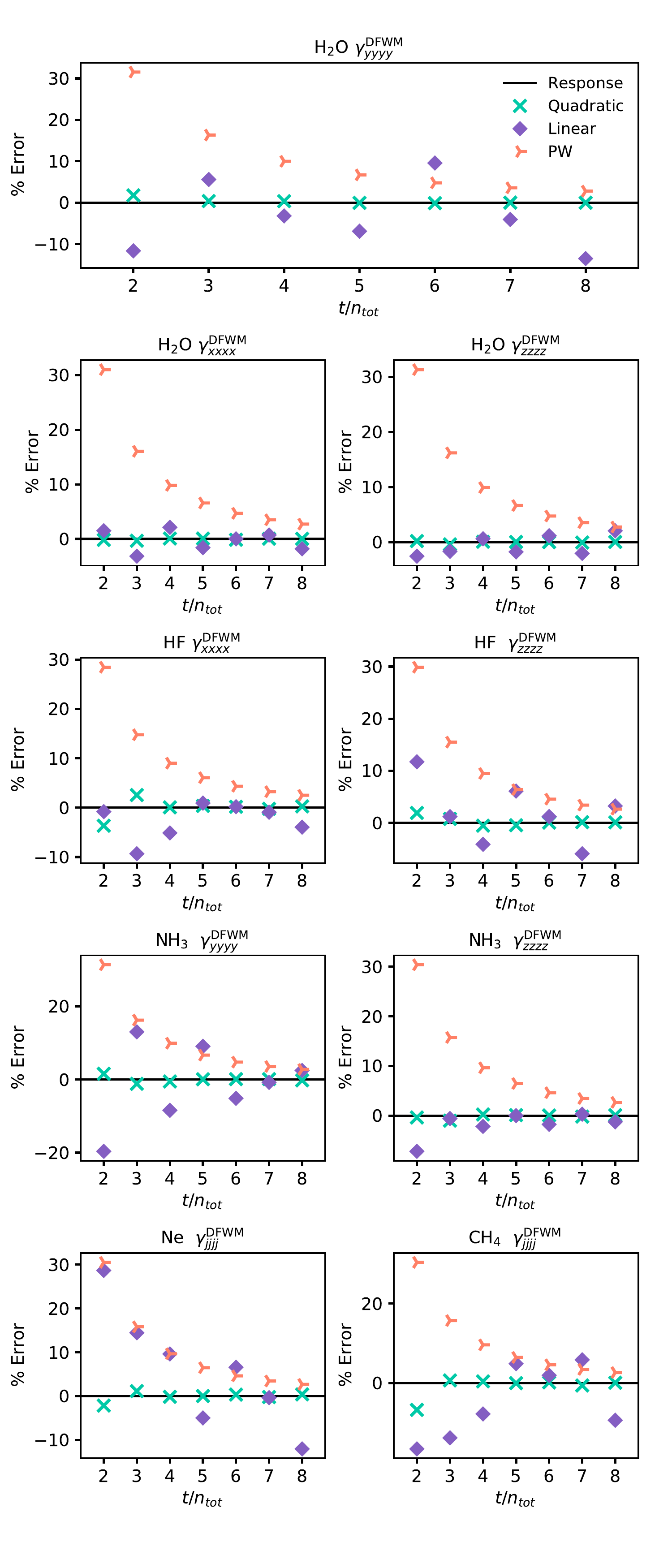} }}
 \subfloat[\centering TDCC2 ]{{\includegraphics[scale=0.55]{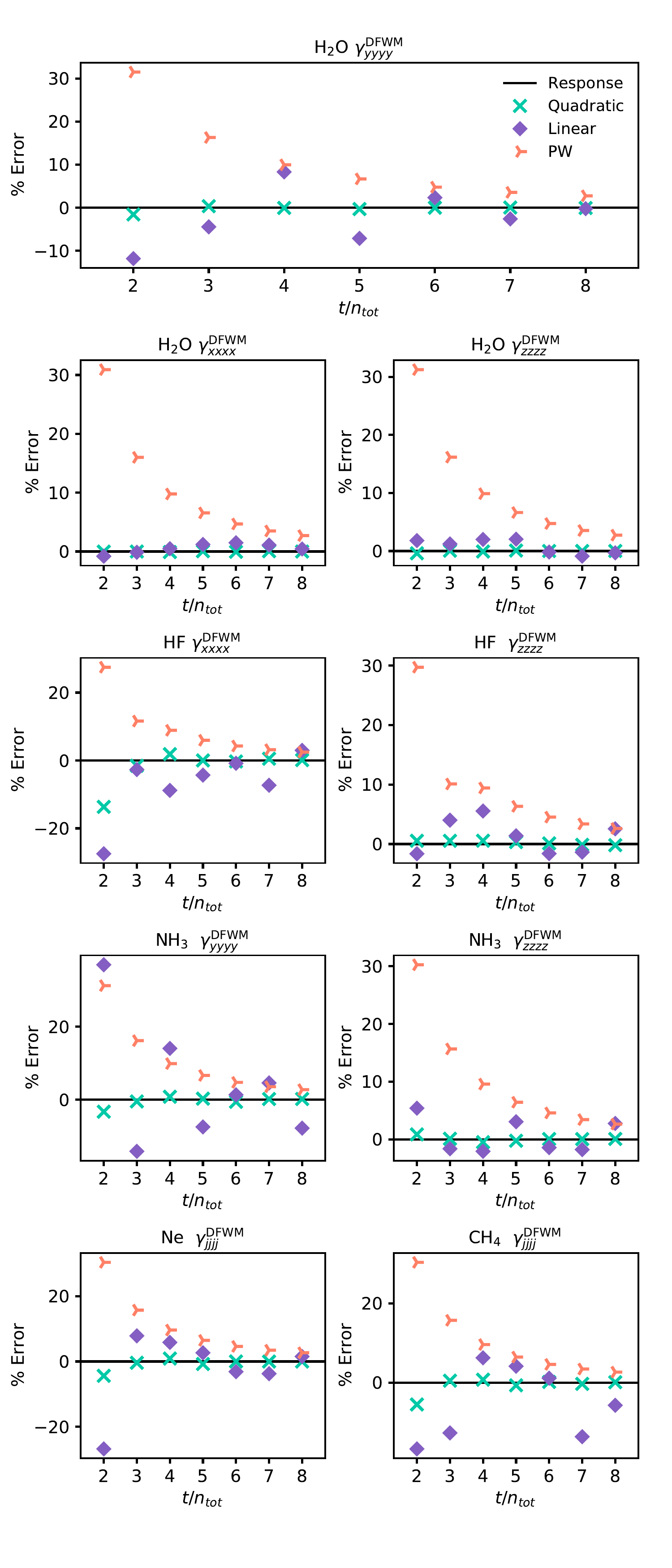} }} \\
\caption{The second hyperpolarizability DFWM \label{fig3}}
\end{center}
\end{figure}

\begin{figure}[H]
\label{second_hyp}
\begin{center}
 \subfloat[\centering TDCCSD ]{{\includegraphics[scale=0.55]{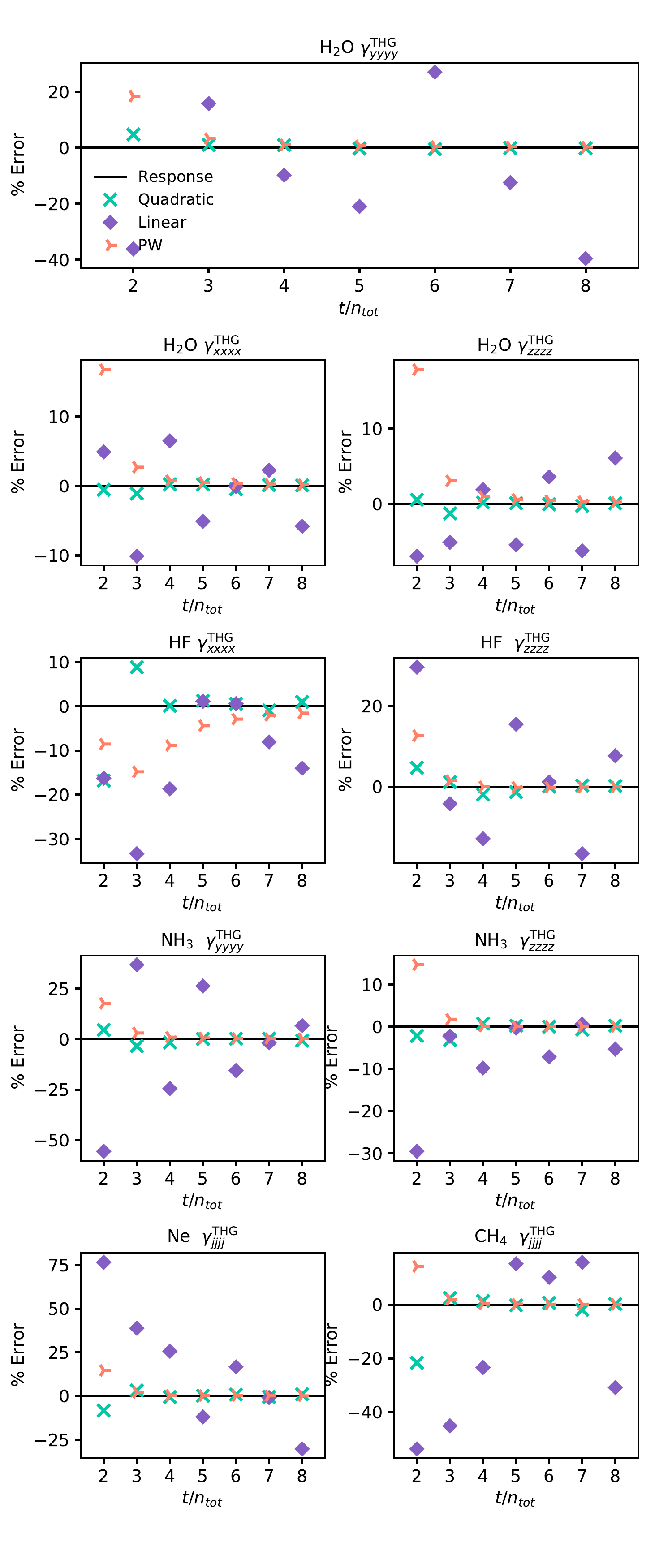} }}
 \subfloat[\centering TDCC2 ]{{\includegraphics[scale=0.55]{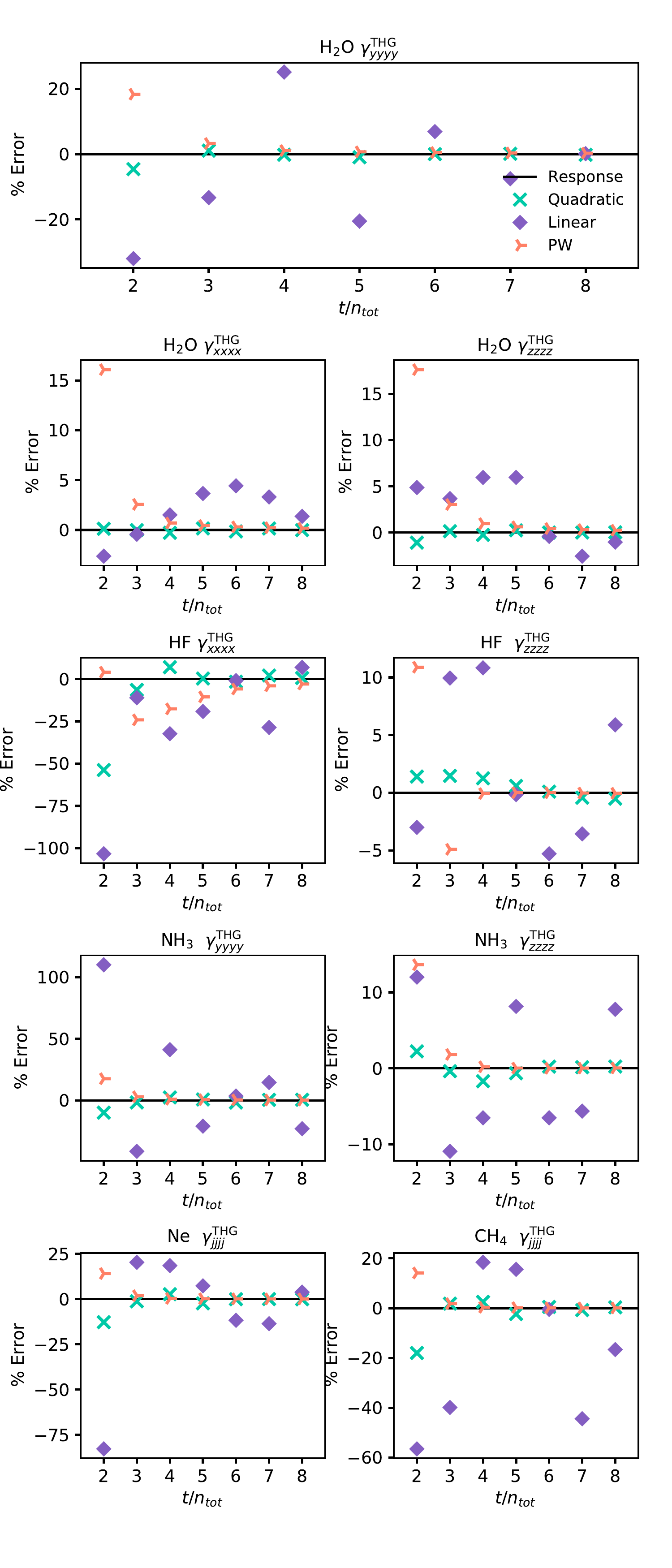} }} \\
\caption{The second hyperpolarizability THG \label{fig4}}
\end{center}
\end{figure}

The full set of figures displaying the accuracy of the extracted property as a function of total simulation time are presented. All unique diagonal Cartesian components of the polarizabilites using the TDCC2 and TDCCSD method are presented in Fig.~\ref{fig1}, the unique SHG and OR components of the first hyperpolarizabilites are presented in Fig.~\ref{fig2}. The DFWM component of the second hyperpolarizablity is presented in Fig.~\ref{fig3} and the THG component of the second hyperpolarizablity is presented in Fig.~\ref{fig4}. 

The properties generated using the TDOMP2 method are not plotted due to lack of available response data, and the properties generated using the TDCIS method are not plotted due to lacking electron correlation which, in turn, renders the method unsuitable. The trends in the figures presented here are overall consistent with those presented and discussed in the main article. 

